\documentclass[aps,pre,epsf,superscriptaddress,amsmath,amssymb,amsfonts,twocolumn,showpacs]{revtex4-1}

\usepackage{graphicx}
\usepackage{epsfig}
\usepackage{dcolumn}
\usepackage{bm}
\usepackage{braket}
\usepackage{amsmath}
\usepackage[english]{babel}
\usepackage{verbatim}
\usepackage{color}

\newcommand{\abs}[1]{\left| #1 \right|}

\begin{document}
\title{Quench Dynamics of Finite Bosonic Ensembles in\\ Optical Lattices with Spatially Modulated Interactions}

\author{T. Pla{\ss}mann}
\affiliation{Zentrum f\"{u}r Optische Quantentechnologien,
Universit\"{a}t Hamburg, Luruper Chaussee 149, 22761 Hamburg,
Germany}  
\author{S. I. Mistakidis}
\affiliation{Zentrum f\"{u}r Optische Quantentechnologien,
Universit\"{a}t Hamburg, Luruper Chaussee 149, 22761 Hamburg,
Germany}  
\author{P. Schmelcher}
\affiliation{Zentrum f\"{u}r Optische Quantentechnologien,
Universit\"{a}t Hamburg, Luruper Chaussee 149, 22761 Hamburg,
Germany}\affiliation{The Hamburg Centre for Ultrafast Imaging,
Universit\"{a}t Hamburg, Luruper Chaussee 149, 22761 Hamburg,
Germany}

\date{\today}

\begin{abstract} 

The nonequilibrium quantum dynamics of few boson ensembles which experience a spatially modulated 
interaction strength and are confined in finite optical lattices is investigated. 
We utilize a cosinusoidal spatially modulated effective interaction strength which is characterized by its wavevector,   
inhomogeneity amplitude, interaction offset and a phase. 
Performing quenches either on the wavevector or the phase of the interaction 
profile an enhanced imbalance of the interatomic repulsion between distinct spatial regions of the lattice is induced. 
Following both quench protocols triggers various tunneling channels and a rich excitation dynamics consisting 
of a breathing and a cradle mode. 
All modes are shown to be amplified for increasing inhomogeneity amplitude of the interaction strength.  
Especially the phase quench induces a directional transport enabling us to discern energetically, otherwise, 
degenerate tunneling pathways. 
Moreover, a periodic population transfer between distinct momenta for quenches of 
increasing wavevector is observed, while a directed occupation of higher momenta can be achieved following a phase quench.  
Finally, during the evolution regions of partial coherence are revealed between the predominantly occupied wells.

\end{abstract}

\maketitle

\section{introduction} 
\label{S1} 

Ultracold atoms in optical lattices have emerged as powerful quantum many-body platforms with 
highly tunable parameters enabling us to emulate in the laboratory a multitude of complex systems \cite{Lewenstein,Yukalov,Lewenstein1}. 
Due to the remarkable experimental progress it is nowadays possible to create arbitrarily shaped potential landscapes \cite{Bloch}, and to  
realize besides many-body also highly controllable few-body systems \cite{Zurn,Serwane,Wenz,Kaufman}. 
Moreover the advent of magnetic \cite{Inouye,Chin_Fesh} and optical Feshbach 
resonances \cite{OFR1,OFR2,Bauer,Lettner,OFRpulses,Yan,OFR4} offer the possibility of tuning the elastic interatomic 
interaction strength with unprecedented level of accuracy. 
In particular optical Feshbach resonances, utilizing optical coupling between bound 
and scattering states, provide the flexibility to design spatially inhomogeneous   
interaction strengths across the atomic sample. 
The corresponding intensity and detuning of the participating optical fields  
can be rapidly changed and allow even for nanometer scale modulations 
of the resulting scattering length \cite{OFRpulses}. 

Spatially inhomogeneous interaction patterns introduce in the system a periodic structure which is 
known as nonlinear optical lattice \cite{Ali,Nicolin,Sekh,Kartashov}. 
This concept reinvogorated the theoretical interest of diverse topics 
ranging from the simulation of sonic black holes \cite{Carusotto,Garay} to altered properties of the emerging 
nonlinear excitations \cite{Kartashov,Kevrekidis_stab}.   
In this latter context a plethora of new phenomena have been revealed such as 
emission of solitons or trains thereof \cite{Rodas,Tsitoura},  
Bloch oscillations of solitary waves \cite{Hang,Salerno,Theocharis1},   
adiabatic compression \cite{Abdullaev1,Theocharis1} and dynamical trapping \cite{Theocharis2} of matter waves 
to name a few. 
Moreover, the existence of a delocalizing transition of bosons in one-dimensional optical lattices \cite{Bludov}, optimal control schemes to stimulate transitions 
into excited modes of a condensate \cite{Hocker}, a particle localization phenomenon at the regions where 
the scattering length vanishes \cite{localization,localization_binary} and the emergence of Faraday waves \cite{Faraday}  
have been demonstrated. 

The above-mentioned investigations 
\cite{Ali,Nicolin,Sekh,Kartashov,Carusotto,Garay,Kevrekidis_stab,Rodas,Tsitoura,Abdullaev1,Theocharis1,Theocharis2,Hang,Salerno,Bludov,localization,localization_binary,Faraday} 
have been performed within the mean-field realm 
resting under the premise of a macroscopic wavefunction which is composed of a single orbital. 
Meanwhile there is evidence that when considering long-range dipolar interactions in bosonic systems fragmentation,  
namely the occupation of more than a single-particle state, occurs \cite{Streltsov_spat,Fischer,xontros,xontros1,xontros2}. 
In this context and referring to few boson ensembles confined in optical lattices different resonant interband 
tunneling mechanisms \cite{Chatterjee2,Chatterjee} have been unveiled. 
Independently and following a linear or a sudden homogeneous interaction quench in lattice trapped few boson systems it has been 
shown that the quench inevitably leads to the population of higher lying band states \cite{Mistakidis6}, it generates collective modes such as 
the breathing and the cradle processes \cite{Mistakidis,Mistakidis1} and couples the lowest 
and excited band states \cite{Mistakidis7,Mistakidis,Jannis}. 
For a lattice setting contact interactions with a spatially varying interaction strength (nonlinear lattice) can give rise to a preferred interaction imbalance of the 
bosons between the distinct lattice sites and therefore particular ground state particle configurations can be formed \cite{Gimperlein}. 
For instance, an intriguing prospect here is to achieve Mott-like few-body states for systems characterized 
by incommensurate filling factors. 
Furthermore, it is particularly interesting to examine whether a certain particle distribution can be displaced in a controlled way upon an interaction quench  
leading to a steered tunneling within the same or energetically different bands during the evolution. 
In this latter context, it is also important to investigate the nature of the quench generated collective modes such as e.g. the cradle mode. 
To address these questions in the present work we employ the Multi-Configurational Time-Dependent Hatree method for 
Bosons (MCTDHB) \cite{MCTDHB1,MCTDHB2} being a multimode treatment which enables us to capture the important correlation effects and 
account for several energetically distinct single-particle bands \cite{Dutta}. 
In this way we investigate, for the first time, the quench induced few boson correlated dynamical response 
in a combined linear and nonlinear optical lattice.    
Our spatially modulated interaction strength is of cosinusoidal form being characterized by its wavevector,   
inhomogeneity amplitude, interaction offset and a phase.     

Regarding the ground state of the system we show that by tuning either the wavevector or the phase,  
the density distribution can be effectively displaced to regions of decreasing interaction strength. 
In particular, for distinct wavevectors the ensemble remains superfluid while a phase shift leads to a displacement 
of the particles in a preferred direction enabling for the existence of Mott-like states. 
The corresponding system's dynamical response upon quenching either the wavevector or the phase of the spatial 
interaction strength is enhanced 
for quenches that yield a non-negligible interaction imbalance of bosons located in different wells. 
Both quench scenarios yield the excitation of a multitude of lowest band interwell tunneling modes 
composed of single-particle and atom pair \cite{Folling,Meinert,Chen} transport. 
The manipulation of these modes by adjusting the interaction offset or 
the inhomogeneity amplitude will also be analyzed and discussed. 
Importantly, by performing a phase quench a directed tunneling along the finite lattice is achieved. 
The latter allows to discriminate between the parity symmetric tunneling modes, e.g. 
single-particle lowest band tunneling from the middle to the left or the right well, which would be otherwise energetically equal. 
Besides the lowest band tunneling dynamics both quenches give rise to an over-barrier transport (being significantly increased when following a phase quench) 
which in turn generates a cradle mode in the outer wells and a global breathing motion of the bosonic cloud.   
These modes are related to single-particle interband processes \cite{Cao_inter,Mistakidis6} to the first, second and fourth excited band respectively, 
and are found to be enhanced for increasing inhomogeneity magnitude.  
Inspecting the one-body momentum distribution a periodic (consecutive) population transfer to higher momenta during the dynamics occurs   
when quenching the wavevector (phase) of the spatially inhomogeneous interaction profile.   
Finally the one-body coherence function reveals a partial coherence between the predominantly occupied wells during the evolution. 

This work is structured as follows.  
In Sec. \ref{S2} we introduce the employed spatially-dependent interaction strength and the multiband expansion which we use for the interpretation of 
the quench induced dynamics. 
Sec. \ref{S3} presents briefly the ground state properties of a system composed of four inhomogeneously interacting bosons in a triple well.  
Then, we focus on the resulting dynamics caused by a quench of the wavevector (Sec. \ref{S4}) or 
the phase (Sec. \ref{S5}) of the spatial interaction strength. 
We provide an outlook and discuss future perspectives in Sec. \ref{conclusion}. 
In Appendix A we discuss the quench induced dynamics for a five-well lattice 
system of filling larger than unity.  
Finally, Appendix B describes our computational approach.

\section{Setup and multiband expansion} 
\label{S2}

The many-body Hamiltonian of $N$ identical bosons possessing mass $M$ and confined in a one-dimensional 
m-well optical lattice reads
\begin{equation}
\begin{split}
H=\sum_{i=1}^N (\frac{p_i^2}{2M} &+ V_0\sin^2(k_{0}x_i))\\&+ \sum_{i<j} V_{int}(x_i-x_j,g,a,k_1,\phi). \label{Hamiltonian}
\end{split}
\end{equation}
The external potential is characterized by its barrier depth $V_0$ and wavevector $k_{0}=\pi/l$, where $l$ 
denotes the distance between successive potential minima. 
In a corresponding experimental setup $k_0$ is the wavevector of the counterpropagating laser beams that form 
the confining optical lattice. 

\begin{figure}[ht]  
  \centering
\includegraphics[width=0.45\textwidth]{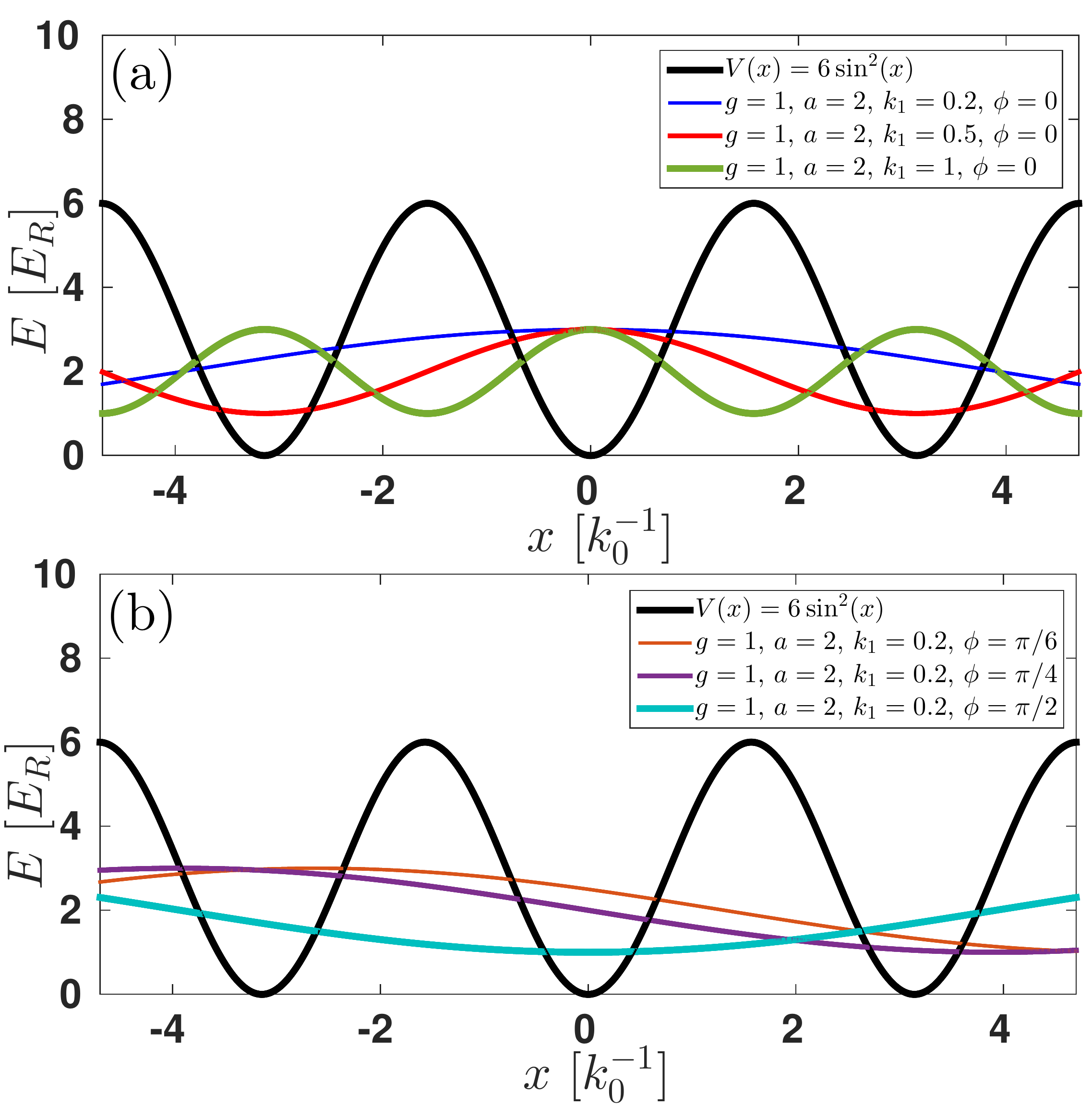}
  \caption{ (a), (b) Different configurations of the spatially-dependent interaction strength $C_{int}(x,g,a,k_1,\phi)$ 
  in a triple well potential $V_0\sin^2(k_0x)$ for $k_0=1$, $V_0=6$ (see legends).} 
  \label{fig:interactionconfiguration}
\end{figure}

The interatomic interaction is modelled by a spatially-dependent short-ranged contact pseudopotential 
$V_{int}\;(x_i-x_j,g,a,k_1,\phi)=C_{int}\;(g,a,k_1,\phi,x_i)\delta(x_i-x_j)$ between particles located 
at positions $x_i$, $i=1,2,...N$. 
The effective one-dimensional spatially-dependent interaction strength reads   
\begin{equation}
C_{int}\;(x,g,a,k_1,\phi)=g\;[\ 1+a\;\cos^2(k_1x+\phi)\ ],   
\label{Eq:interactioncoefficient}
\end{equation} 
where $g$ refers to an average interaction offset. 
$k_1$ denotes the wavevector of the  
periodic modulation, $a$ is the amplitude of the inherent inhomogeneity and $\phi$ is a constant phase shift.  
Note that $\phi\neq0$ and fixed $k_1$ yields an interaction strength imbalance between all lattice wells, 
while for varying $k_1$ and $\phi=0$ $C_{int}$ is on average the same only for the parity 
symmetric, with respect to the center ($x=0$), outer sites. 
Due to periodicity $\phi$ takes values within the interval $[0,\pi/2]$.  
Several interaction profiles of Eq. (\ref{Eq:interactioncoefficient}) for varying wavevector $k_1$ or 
phase $\phi$ are presented in Fig. \ref{fig:interactionconfiguration} together with the underlying triple well potential $V_0\sin^2(k_0x)$. 
Experimentally such a spatially modulated interaction profile can be achieved with the aid of optically 
induced Feshbach resonances \cite{OFR1,OFR2,OFR5,OFR4}, e.g. by a laser field tuned near a 
photoassociation transition. 
Alternatively a technique of holographic beam shaping can be used, e.g. a digital micromirror device \cite{DMD}, 
to engineer wavefronts of arbitrary phase, amplitude and wavelength \cite{DMD1,DMD2}. 
 
For simplicity, the Hamiltonian is rescaled in units of the recoil energy $E_R=\frac{\hbar^2k_{0}^2}{2M}$. 
Thus the frequency, spatial and temporal scales are given in units of $\omega_R$, $k_{0}^{-1}$ and 
$\hbar E_R^{-1}$ respectively.    
In addition we set $\hbar=M=k_{0}=1$.  
The confinement of the bosons in the $m$-well system is ensured by the use of hard-wall boundary 
conditions at $x_m=\pm \frac{m\pi}{2k_{0}}$. 
Finally, the lattice depth is fixed to $V_0=6E_R$ including this way two localized 
single-particle Wannier states per lattice site.  

To examine the static properties and the quench induced dynamics upon varying the wavevector $k_1$ or the 
phase $\phi$ of $C_{int}$ we employ MCTDHB \cite{MCTDHB1,MCTDHB2}.  
In contrast to the mean-field approximation, within this approach we exploit an expansion in terms of many  
variationally optimized time-dependent single-particle functions (for more details, see also Appendix A).  
The latter allows for the investigation of the emergent interparticle correlations revealing the many-body properties of the system.  
To identify the modes participating in the dynamics we project the numerically obtained many-body correlated MCTDHB wavefunction on a 
time-independent number state basis consisting of single-particle Wannier states being localized on each lattice site. 
Such an expansion offers the possibility to study inter- and intraband transitions \cite{Mistakidis}.  
The many-body bosonic wavefunction of $N$ bosons in an $m$-well potential which includes $j$ localized Wannier states reads
\begin{equation}
|\Psi(t)\rangle=\sum_{\vec{n}} C_{\vec{n}}(t)|\vec{n}\rangle.\label{multiband}
\end{equation} 
The multiband Wannier number state $|\vec{n}\rangle=|\otimes_{\lambda=0}^{j-1}n_1^{(\lambda)},..., \otimes^{j-1}_{\lambda=0}n_m^{(\lambda)}\rangle$ while   
the Wannier occupation number $n_i^{(\lambda)}$ indicates the number of bosons that reside in the Wannier orbital 
$|n_i^{(\lambda)}\rangle$ of the $i$-th well and $\lambda$-th energy band. 
Due to the fixed number of bosons $N$ the total number of configurations is constrained by $\sum_{i=1}^{m}\sum_{\lambda=1}^{j-1}n_i^{(\lambda)}=N$.   
For a setup of $N=4$ bosons confined in a triple well $m=3$, which will be our workhorse in the following, e.g.  
the state $|1^{(0)},1^{(1)}\otimes1^{(1)},1^{(0)}\rangle$ indicates that in the left and right wells one boson occupies the Wannier orbital of the 
energetically lowest band while the remaining two atoms are in the middle well, residing in the Wannier orbital of the first excited band. 
For simplicity, below, we shall omit the zero index when referring to the energetically lowest (zeroth) band. 

We note that in the case of a homogeneous contact interaction and regarding the zeroth band states one can 
realize four distinct energetic classes of number states.   
Namely, the single pairs (SP) $\{|2,1,1\rangle +\circlearrowleft \}$, double pairs (DP) $\{|2,0,2 \rangle +\circlearrowleft\}$, 
triples (T) $\{|3,1,0 \rangle + \circlearrowleft \}$ and quadruples (Q) $\{|4,0,0 \rangle + \circlearrowleft \}$,  
where $\circlearrowleft$ denotes all corresponding permutations.  
However, in the presence of a spatially inhomogeneous interaction, each energy class is further energetically splitted  
depending on the combination of the corresponding occupation number $n_i$ and the spatially averaged interaction strength in the $i$-th well.   
Here we distinguish two cases. 
For varying $k_1$ and $\phi=0$ each energy class splits into a subclass containing the states 
with the lowest occupancy in the middle well and another one which includes all the other states of the original energy class. 
As an example the SP class separates into the $\{|2,1,1\rangle, |1,1,2\rangle  \}$ and $\{|1,2,1\rangle \}$ subclasses. 
Moreover, since the phase shift $\phi$ yields distinct $C_{int}$ in each well all states of a certain 
class become energetically individual. 

\subsection{Basic Analysis Tools}

We next briefly introduce the main observables that will be employed for the interpretation of 
the quench induced nonequilibrium dynamics on both the one- and two-body level.   

Performing a quench we change abruptly a parameter $\zeta$ of the system [here, the wavevector $k_1$ 
or the phase $\phi$ of $C_{int}$, see also Eq. (\ref{Eq:interactioncoefficient})] from an initial value 
$\zeta_0 = \zeta(t=0)$ to a final one $\zeta_f$. 
Then, the ground state $\left| {\Psi(0)}\right\rangle$ of the initial Hamiltonian $H(\zeta_0)$ 
evolves at time $t$ according to $\ket{\Psi(t)}\equiv\left|\Psi_{\zeta_f} \right\rangle =U_{\zeta_f}\left| {\Psi(0)} \right\rangle = \exp(-iH_{\zeta_f}t/\hbar)\left| {\Psi(0)} \right\rangle$ 
under the influence of the $\zeta_f$-quenched Hamiltonian. 
The corresponding overlap between the initial (ground) and the time-evolving wavefunction \cite{Gorin,Venuti,Campbell} 
yields the fidelity of the system which reads 
\begin{equation}
F(t;\zeta_f)=|\langle \Psi(0)|\Psi(t;\zeta_f) \rangle|^2. \label{Fid}
\end{equation}
This quantity is a time-resolved measure for the effect of the quench onto the system and therefore 
includes information about its dynamical response following the quench \cite{Mistakidis,Mistakidis1,Mistakidis4,Mistakidis6,Mistakidis7,Jannis,Campbell}. 

To identify the degree of one-body correlations during the quench dynamics, we utilize the first order 
coherence function \cite{Naraschewski,Sakmann_cor} 
\begin{equation}
g^{(1)}(x,x')=\frac{\rho^{(1)}(x,x')}{\sqrt{\rho^{(1)}(x)\rho^{(1)}(x')}}. \label{one_body_correlation}
\end{equation} 
The one-body reduced density matrix $\rho^{(1)}(x,x')=\langle x|\hat{\rho}^{(1)}|x'\rangle$ is obtained by tracing out 
all bosons but one in the $N$-body density operator $\hat{\rho}^{(N)}=|\Psi\rangle\langle\Psi|$ of the $N$-body system. 
$|g^{(1)}(x,x')|$ takes values within the interval $[0,1]$. 
Two distinct spatial regions $\mathcal{D}$, $\mathcal{D}'$, with $ \mathcal{D} \cap \mathcal{D}' = \varnothing$, where    
$|g^{(1)}(x,x';t)|= 0$, $x\in \mathcal{D}$, $x'\in \mathcal{D}'$ are said to be fully incoherent, while if 
$|g^{(1)}(x,x';t)|= 1$, $x\in \mathcal{D}$, $x'\in \mathcal{D}'$ holds these are termed perfectly coherent. 
For bosonic ensembles in optical lattices it is known that if within a well $|g^{(1)}(x,x';t)|= 1$ (diagonal elements)  
while between different wells $0\ll|g^{(1)}(x,x';t)|\leq1$ [$|g^{(1)}(x,x';t)|= 0$] (off-diagonals) the appearance of  
superfluid-like [Mott-like] one-body correlations are indicated. 
$g^{(1)}(x,x')$ measures the deviation of the many-body state from a mean-field product state for a given set of spatial 
coordinates $x$, $x'$. 
Then the absence of one-body correlations is indicated by $|g^{(1)}(x,x';t)|=1$ for every $x$, $x'$ while if at 
least two distinct spatial regions are partially incoherent, i.e. $|g^{(1)}(x,x';t)|<1$, the emergence of 
one-body correlations is signified. 

To infer about the degree of second order correlations, we resort to the 
two-body coherence function \cite{Sakmann_cor} 
\begin{equation}
g^{(2)}(x_1,x_2)=\frac{\rho^{(2)}(x_1,x_2)}{\rho^{(1)}(x_1)\rho^{(1)}(x_2)}. \label{two_body_correlation} 
\end{equation} 
The two-body density $\rho^{(2)}(x_1,x_2)=\langle x_1 x_2|\hat{\rho}^{(2)}|x_1 x_2\rangle$ is obtained by 
a partial trace over all but two bosons of the $N$-body density operator. 
It refers to the probability of finding two atoms located at positions $x_1$, $x_2$ at time $t$. 
A many-body state characterized by $|g^{(2)}(x_1,x_2)|=1$ is termed fully second order coherent 
or uncorrelated, while if $|g^{(2)}(x_1,x_2)|$ takes values larger (smaller) than unity it is referred to as 
correlated (anti-correlated). 

In order to visualize the spatially resolved system dynamics we study 
\begin{equation}
\delta \rho^{(1)}(x,t)=\rho^{(1)}(x,t)-\langle\rho^{(1)}(x)\rangle_T. \label{fluctuations}
\end{equation}
It refers to the deviation of the one-body density from its time average 
$\langle\rho^{(1)}(x)\rangle_T=\int_0^T  \rho^{(1)}(x,t)/T$ over the considered propagation time $T$.  
In this sense, $\delta\rho^{(1)}(x,t)$ encompasses the temporal fluctuations of the one-body density 
around its mean along the finite lattice \cite{Mistakidis,Mistakidis1}. 

The collective expansion and contraction (breathing) dynamics \cite{Abraham_breath,Abraham_breath1} of the bosonic cloud, within a spatial region $\mathcal{D}$   
of the lattice, can be measured via the position variance 
\begin{equation}
\sigma^2_{\mathcal{D}}(t)=\braket{\Psi(t)|\hat{x}_{\mathcal{D}}^2|\Psi(t)}-\braket{\Psi(t)|\hat{x}_{\mathcal{D}}|\Psi(t)}^2. \label{variance} 
\end{equation}
Here, the one-body operators correspond to $\hat{x}_{\mathcal{D}}=\int_{-d}^{d} dx x \hat{\Psi}^{\dagger}(x)\hat{\Psi}(x)$ and 
$\hat{x}_{\mathcal{D}}^2=\int_{-d}^{d} dx x^2 \hat{\Psi}^{\dagger}(x) \hat{\Psi}(x)$. 
$\hat{\Psi}(x)$ [$\hat{\Psi}^{\dagger}(x)$] is the field operator that annihilates [creates] a boson at position $x$ and $-d$, $d$ refer to the edges 
of the spatial region $\mathcal{D}$ under consideration. 
The position variance, evaluated over the entire lattice, essentially quantifies 
a ’global breathing’ mode consisting of interwell tunneling and intrawell breathing modes \cite{Koutentakis}. 
We remark that in our case, see also the discussion in Secs. \ref{S4_tunneling} and \ref{S5B}, the breathing mode predominantly occurs 
within the middle well of the finite lattice. 
For this reason we shall calculate the position variance in the middle well, denoted by the index $M$, namely 
$\sigma^2_M(t)$ with $\mathcal{D} \equiv [-\pi/2,\pi/2]$ i.e. $d=\pi/2$ in Eq. (\ref{variance}). 

Finally, we inspect the momentum distribution of the one-body reduced density matrix $\rho^{(1)}(x,x';t)$  
\begin{equation}
n(k,t)=\frac{1}{2\pi} \int \int dx dx' \rho^{(1)}(x,x',t)e^{-ik(x-x')t},\label{momentum}   
\end{equation} 
with the aim of understanding whether a certain multitude of momenta is populated during the dynamics 
as a consequence of the employed quench protocol. 
This quantity is a routinely employed observable in quantum gas experiments, 
since it is accessible via time-of-flight measurements \cite{Bloch}.

\section{Ground state properties}
\label{S3}

Before investigating the dynamics, let us elaborate on the ground state properties of a system with filling $\nu=N/m>1$ (here $N=4$ and $m=3$) under the influence 
of the spatially-dependent interaction profile [see Eq. (\ref{Eq:interactioncoefficient})].  
Referring to the case of a homogeneous contact interaction the ground state characteristics of a lattice setup  
depends strongly on the system's filling factor.  
For commensurate fillings ($\nu=1,2,...$) and increasing interatomic repulsion one can realize the superfluid to Mott insulator
phase transition \cite{Motttheo,Mottexp}, while for incommensurate fillings ($\nu\neq1,2,...$) the delocalized fraction of 
particles forbids the occurrence of a Mott state due to prevailing on-site interaction effects. 
On the other hand, spatially varying interactions can influence mainly systems consisting of sufficiently overlapping atoms (namely $\nu \geq 1$) as 
the emergent on-site interactions effects can be highly exploited in this case. 
In the following we explore the ground state properties for $N=4$ bosons in a triple well either for varying wavevector $k_1$ or phase $\phi$ but for fixed 
interaction offset, $g=1$, and inhomogeneity amplitude $a=2$. 

\begin{figure*}[ht] 
  \centering
\includegraphics[width=1.0\textwidth]{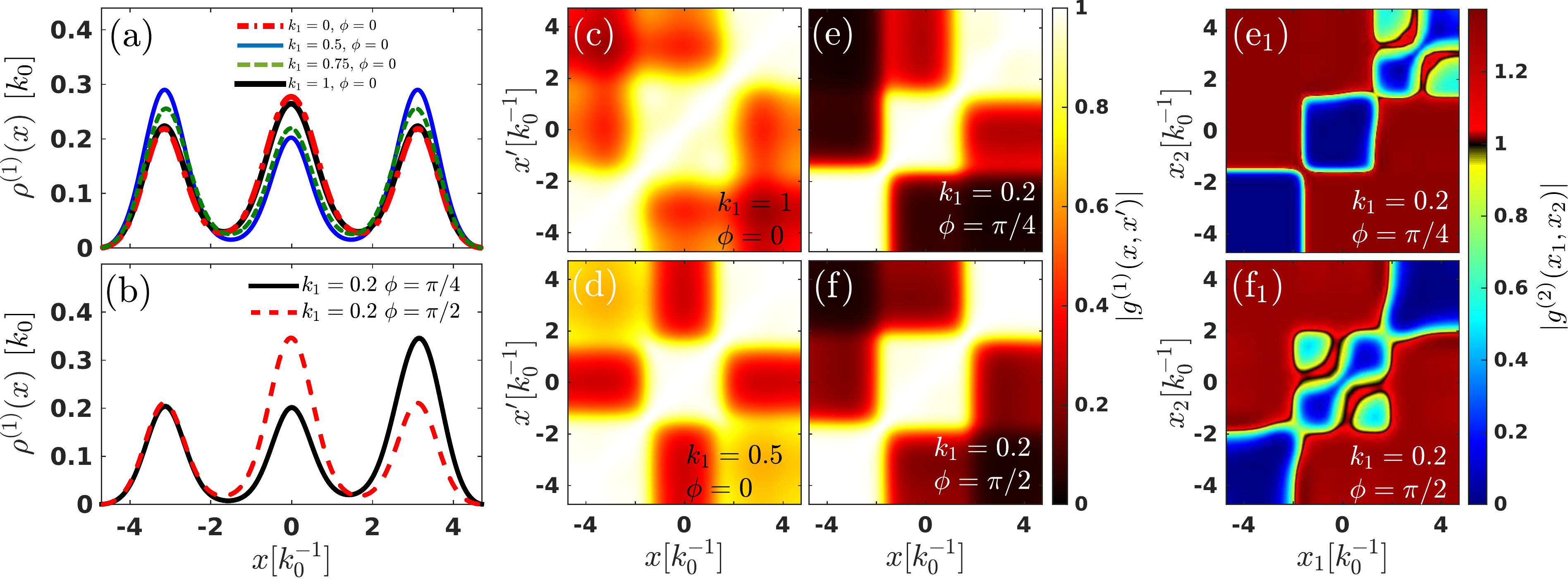}
  \caption{ One-body density $\rho^{(1)}(x)$ for either varying ($a$) wavevector $k_1$ or ($b$) phase $\phi$ of the 
  spatial interaction strength $C_{int}$ (see legends). 
  The remaining parameters that characterize $C_{int}$ are $g=1$ and $a=2$. 
  ($c$)-($f$) The corresponding first order coherence $|g^{(1)}(x,x')|$ for the selected ground states shown in ($a$) and ($b$) (see legends).   
  ($e_1$), ($f_1$) The two-body coherence function $|g^{(2)}(x_1,x_2)|$ for the ground state configurations illustrated in ($e$) and ($f$) respectively. 
  The system consists of four bosons confined in a triple well.} 
  \label{fig:groundstate}
\end{figure*}

We first inspect the dependence of the ground state configuration on the 
wavevector $k_1$ of the interaction strength $C_{int}$ for $\phi=0$. 
Fig. \ref{fig:groundstate} ($a$) shows the one-body density $\rho^{(1)}(x)=\rho^{(1)}(x,x'=x)$ for different spatial periodicities   
$k_1=0,0.5,0.75$ and $1$ of $C_{int}$.  
The various $k_1$ values lead to distinct spatially averaged interaction strengths for the central and the outer wells. 
For $k_1=1$ the bosons residing in each well are subject to the same average interaction strength, see also Fig. \ref{fig:interactionconfiguration}.  
A slightly increased particle density in the middle well when compared to the outer ones is observed due to the hard-wall boundary conditions. 
This situation resembles a homogeneously interacting system i.e. $k_1=0$ with $C_{int}=g+a$, see Fig. \ref{fig:groundstate} ($a$).  
On the contrary, $k_1=0.5$ yields a spatially varying interaction strength exhibiting a peak within the central well [see Fig. \ref{fig:interactionconfiguration}] 
which forces the ensemble to preferably populate the outer wells. 
Other values of $k_1$ yield an intermediate behavior of the particle density, e.g. see $\rho^{(1)}(x)$ for $k_1=0.75$ in Fig. \ref{fig:groundstate} ($a$).   

To cast light on the non-local properties of the system's ground state from a one-particle perspective we employ the first order coherence   
function \cite{defcoherence,Naraschewski}, see also Eq. (\ref{one_body_correlation}).  
In this sense the above-described ground state single-particle density distributions caused by $C_{int}$ for $k_1=1$ and $k_1=0.5$ possess  
superfluid-like one-body correlations, see Figs. \ref{fig:groundstate} ($c$) and ($d$) respectively. 
Therefore we can infer that the ensemble under the influence of different spatial periodicities $k_1$ with all the other parameters of $C_{int}$ kept fixed  
exhibits a superfluid behavior, being an anticipated result since $\nu>1$.   
Moreover, the difference with respect to the magnitude of the off-diagonal $|g^{(1)}|^2$ elements reflects essentially 
the particular density imbalance between the different wells induced by $k_1$, see also Fig. \ref{fig:groundstate} ($a$). 
Namely, the most occupied wells with the larger spatially integrated density exhibit stronger coherence losses 
with the lower populated ones than the latter among each other. 

To infer about the effect that finite phase terms have on the ground state properties of the system we consider in the following the case $\phi\neq0$.  
As it can be easily seen from Eq. (\ref{Eq:interactioncoefficient}) a phase $\phi>0$ accounts for a spatial shift of the  
interaction profile $C_{int}$ yielding an imbalance of the interparticle repulsion between all lattice wells.  
With the aid of such spatially-dependent interaction strengths a directed shift of the single-particle density distribution can be achieved. 
For instance, the choice $k_1=0.2$ and $\phi=\pi/4$ corresponds to a spatial interaction strength with a minimum (maximum) in 
the vicinity of the right (left) well [see also Fig. \ref{fig:interactionconfiguration}]. 
As a result two atoms mainly populate the right well and one resides in each of the left 
and middle wells, see Fig. \ref{fig:groundstate} ($b$). 
Therefore an important contribution to the ground state configuration stems from the number state $\ket{1,1,2}$. 
Moreover, setting $k_1=0.2$ and $\phi=\pi/2$ yields a $C_{int}$ which is minimized around the middle well and maximized 
in each of the outer wells. 
Then, a bunching of the atoms is observed in the central lattice region and the one-body density is 
described by the state $\ket{1,2,1}$, see Fig. \ref{fig:groundstate} ($b$).  
Summarizing, the low values of $C_{int}$ in the neighborhood of either the middle ($\phi=\pi/2$) 
or the right ($\phi=\pi/4$) well gives rise to double occupation in these wells of the lattice and to a single 
occupation in the other wells.  
The possibility to create such almost localized single-particle density distributions by tuning the 
phase $\phi$ of the interaction profile $C_{int}$ 
enables us to emulate spatially inhomogeneous Mott insulator like states.  
The latter can be firstly confirmed by employing the corresponding one-body coherence   
function $|g^{(1)}(x,x')|$, shown in Figs. \ref{fig:groundstate} ($e$) and ($f$).   
Indeed, $|g^{(1)}(x,x')|$ exhibits almost vanishing off-diagonal contributions which 
suggest the emergence of Mott-like correlations. 
To further ensure the existence of the Mott insulator like state we rely on the two-body coherence 
function \cite{Sakmann_cor}, see also Eq. (\ref{two_body_correlation}). 
Fig. \ref{fig:groundstate} ($e_1$) presents $|g^{(2)}(x_1,x_2)|$ for $k_1=0.2$ and $\phi=\pi/4$. 
An anti-correlated behavior occurs in the left and central wells, while all the off-diagonal elements are correlated. 
Most importantly, a correlated behavior takes place within the right well signalled by two correlation holes \cite{Sakmann_cor}, 
see in particular the substructures where $|g^{(2)}(x_1,x_1)|>1$. 
These indicate that two particles are likely to reside in the right well and only one in each of the remaining wells, confirming 
once more the existence of the Mott-like state $\ket{1,1,2}$. 
The same behavior, in terms of $|g^{(2)}(x_1,x_2)|$, is observed for $C_{int}$ with $k_1=0.2$ and $\phi=\pi/2$ but this time 
$|g^{(2)}(x_1,x_2)|>1 (<1)$ in the middle (outer) well [Fig. \ref{fig:groundstate} ($f_1$)] resulting in the Mott-like state $\ket{1,2,1}$. 
Note that particle localization within regions of a vanishing scattering length has already been reported for inhomogeneously 
interacting bosons in a box potential \cite{localization}.   

Concluding, a spatially-dependent interaction strength enables for the emergence of Mott-like correlations (besides the inherent 
superfluid character due to $\nu>1$) or even the possibility to shift the particles to a preferred direction. 
Taking advantage of the different spatial interaction profiles offers the opportunity to prepare certain ground state configurations.  
For instance, concerning larger lattice systems a sequence of inhomogeneous Mott-like states such as a double occupation for every 
second well can be achieved.

\section{Dynamics Following a Quench of the Period of the Interaction Strength} 
\label{S4} 

In the present section the nonequilibrium dynamics upon a sudden change of the wavevector $k_1$ of the spatially-dependent interaction 
strength $C_{int}$ is examined. 
The system consists of four bosons confined in a triple well and it is initialized in the ground state of the many-body Hamiltonian 
given by Eq. (\ref{Hamiltonian}) with a spatial interaction strength coefficient $C_{int}$ [Eq. (\ref{Eq:interactioncoefficient})]  
characterized by the parameters $g=1$, $k_1=0$, $a=2$ and $\phi=0$. 
Then initially ($t<0$) $C_{int}=3$ and to induce the dynamics we consider a quench of the wavevector of $C_{int}$. 
Specifically the dynamics is governed by the Hamiltonian of Eq. (\ref{Hamiltonian}) with $C_{int}= 1+2\;\cos^2(k_1x)$. 
The initial many-body state is an admixture of the number states $|1,2,1\rangle$,   
$|1,1,2\rangle$, $|2,1,1\rangle$, and $|1,3,0\rangle$, $|0,3,1\rangle$ with 
the $|1,2,1\rangle$, $|1,3,0\rangle$ and $|0,3,1\rangle$ possessing the 
dominant contribution due to the hard-wall boundary conditions. 

\subsection{Tunneling properties} 
\label{S4_tunneling}

\begin{figure}[ht]  
  \centering
\includegraphics[width=0.5\textwidth]{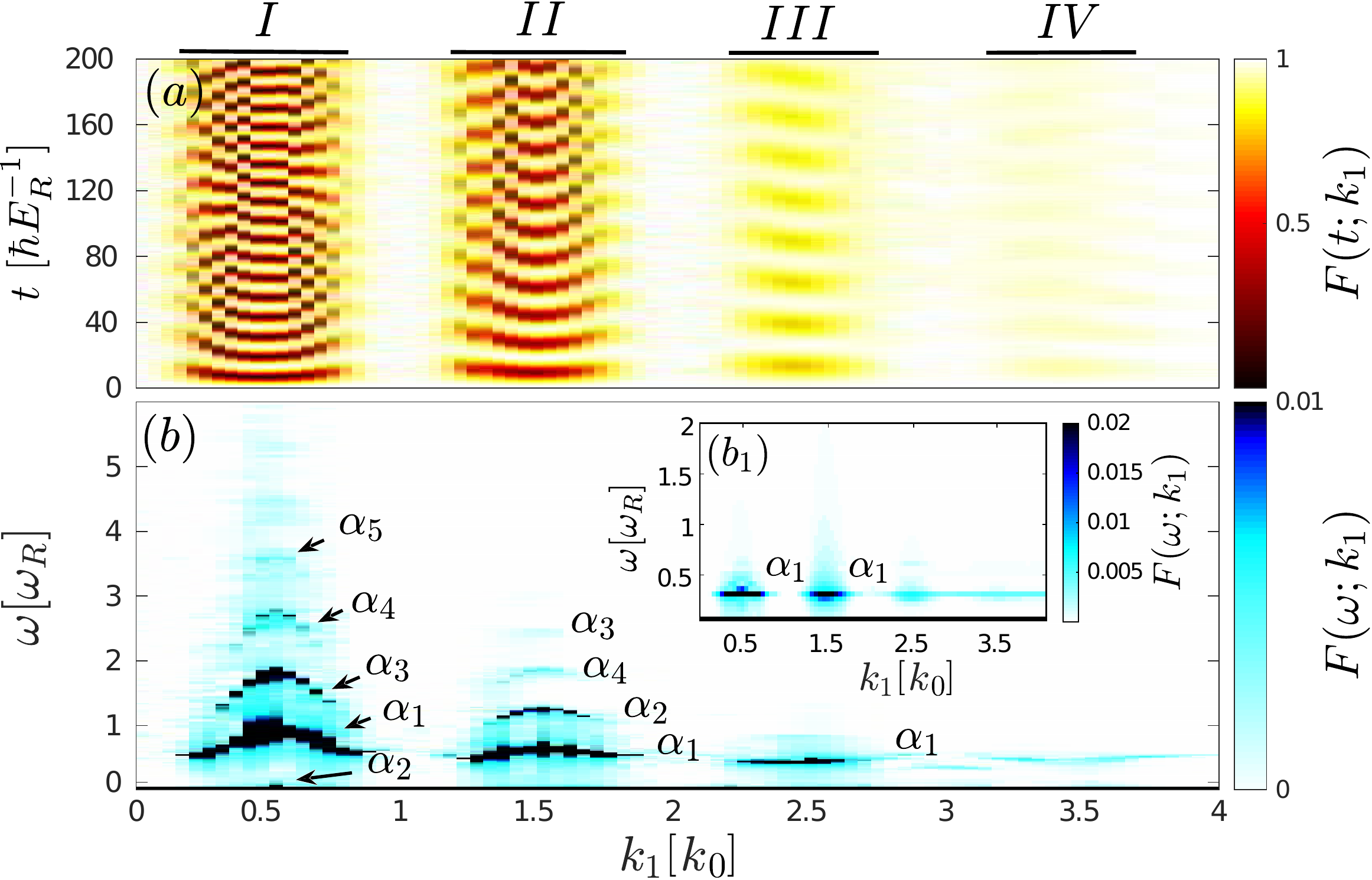}
  \caption{ ($a$) Fidelity evolution $F(t;k_1)$ for varying wavevector $k_1$ of the spatially-dependent interaction strength $C_{int}$ for $g=1$, $a=2$, $\phi=0$.    
  ($b$) The corresponding spectrum $F(\omega;k_1)$.  
  Inset ($b_1$) presents $F(\omega;k_1)$ for $g=0.1$, while all other system parameters are the same as in ($a$).  
  For $t=0$ we choose the ground state of four bosons in a triple well.}  
  \label{fig:fidelity1}
\end{figure}

To infer about the system's dynamical response upon quenching $k_1$ we first rely on the fidelity evolution 
$F(t;k_1)=|\langle \Psi(0)|\Psi(t;k_1) \rangle|^2$ which provides the overlap between the initial (ground) and the 
time-evolving wavefunction \cite{Gorin,Venuti,Campbell}.  
Fig. \ref{fig:fidelity1} ($a$) illustrates $F(t;k_1)$ for varying wavevector $k_1$. 
The dynamics is characterized by enhanced response regions ($F(t;k_1)\ll1$) centered around $k_1=d/2$ with $d=1,3,5,7$ denoted 
by $I$, $II$, $III$ and $IV$ respectively in Fig. \ref{fig:fidelity1} ($a$) and 
regions of low response ($F(t;k_1)\approx1$) located around integer values of $k_1$. 
In the former case bosons in different wells are subject to distinct spatially averaged interaction strengths, while in 
the latter case all wells share on average the same interaction strength [see also Eq. (\ref{Eq:interactioncoefficient}) and Fig. \ref{fig:interactionconfiguration}].  
Within the enhanced response regions $F(t;k_1)$ exhibits an oscillatory behavior in time which gradually transforms from a multifrequency pattern for small $k_1$ values 
(e.g. region $I$) to a single frequency one for increasing $k_1$ (e.g. see region $III$ and the discussion below). 
In addition, with increasing $k_1$ the enhanced response regions gradually loose amplitude [e.g. $F(t;k_1\approx3.5)\approx 0.9$] due to the fact that 
the averaged spatially-dependent interaction strength tends to a homogeneous configuration. 

\begin{figure}[ht]  
  \centering
\includegraphics[width=0.5\textwidth]{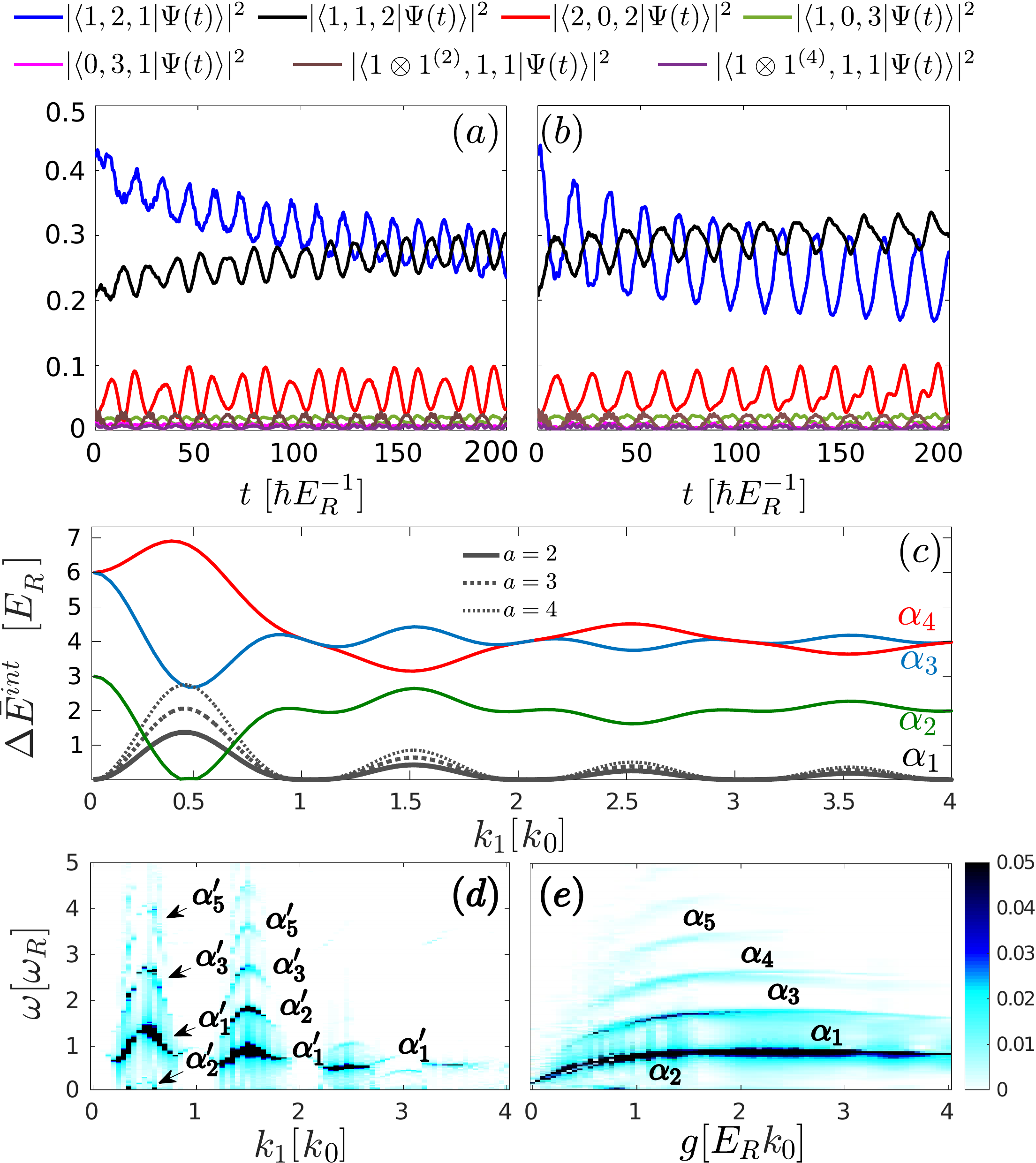}
  \caption{ Time-evolution of different number state probabilities (see legend) upon quenching the wavevector from $k_1=0$ to either 
  ($a$) $k_1=0.2$ or ($b$) $k_1=0.5$. 
  Parameters of the spatially-dependent interaction strength $C_{int}$ are $g=1$, $a=2$ and $\phi=0$. 
  ($c$) Average interaction energy difference $\Delta \bar{E}^{int}$ [see also Eq. (\ref{aver_int})] for the tunneling branches shown in Fig. \ref{fig:fidelity1} ($b$) that 
  correspond to different number state transitions.   
  ($d$) $F(\omega;k_1)$ for $a=5$, $g=1$, $\phi=0$ and varying $k_1$.  
  ($e$) $F(\omega;g)$ for $k_1=0.5$, $a=2$, $\phi=0$ and varying interaction offset $g$. 
  In all cases the system is initially prepared within the ground state of four bosons in a triple well.}  
  \label{fig:fidelity1_a}
\end{figure}

To assign the tunneling modes triggered by a quench of $k_1$ we employ the fidelity spectrum 
$F(\omega;k_1)=\operatorname{Re} \left\{ \frac{1}{\pi} \int dt\;F(t;k_1)e^{i\omega t} \right\}$ \cite{Jannis,Mistakidis7}, see Fig. \ref{fig:fidelity1} ($b$). 
As can be observed, tunneling occurs only within the enhanced response regions and the number of participating modes 
strongly depends on the magnitude of $k_1$. 
Within the region $I$ ($k_1\approx0.5$) five distinct tunneling modes appear corresponding to the frequency branches 
$\alpha_1$-$\alpha_5$ in Fig. \ref{fig:fidelity1} ($b$). 
To identify the corresponding dominant particle configurations $\ket{\vec{n}}$ that are responsible for the occurence of these tunneling branches upon quenching 
the wavevector $k_1$ of $C_{int}$ we utilize the multiband expansion introduced in Eq. (\ref{multiband}). 
Thus, we calculate the number state probabilities, during the evolution, defined as $\abs{\braket{\vec{n}|\Psi(t)}}^2$. 
Figures \ref{fig:fidelity1_a} ($a$) and ($b$) present the time-evolution of $\abs{\braket{\vec{n}|\Psi(t)}}^2$ for the most significantly contributing number states when 
performing a wavevector quench from $k_1=0$ to $k_1=0.2$ and $k_1=0.5$ respectively. 
Most importantly, by calculating the frequency spectrum of $\abs{\braket{\vec{n}|\Psi(t)}}^2$ we are able to relate these transition probabilities to the modes 
manifested in $F(\omega;k_1)$ e.g. for the values of $k_1$ after the quench shown in Figs. \ref{fig:fidelity1_a} ($a$) and ($b$). 
However, the same procedure has been followed for all quench amplitudes to be presented below (not shown here for brevity reasons).  
As a consequence, the aforementioned modes can be energetically categorized as follows.   
The most dominant process (see branch $\alpha_1$) refers to single-particle tunneling e.g. $|1,2,1\rangle \leftrightharpoons |1,1,2\rangle$. 
Interestingly enough the second order tunneling mode $|1,2,1\rangle \leftrightharpoons |2,0,2\rangle$ (see branch $\alpha_2$) occurs  
at smaller frequencies (for an explanation see the discussion below). 
Moreover, the frequency branches $\alpha_3$ and $\alpha_4$ correspond to transitions between the 
SP and T modes and in particular to e.g. $|1,1,2\rangle \leftrightharpoons |1,0,3\rangle$ (single-particle tunneling) 
and $|0,3,1\rangle \leftrightharpoons |1,1,2\rangle$ (atom pair tunneling \cite{Folling,Meinert,Chen}) respectively.  
Finally, the highest frequency mode $\alpha_5$ refers to an interband transition and will be addressed in the next subsection.  
Turning to region $II$ we observe the occurrence of the same tunneling modes as in $I$ but overall shifted to smaller frequencies 
while the interband mode $\alpha_5$ dissapears.  
More importantly, the second order tunneling mode indicated by $\alpha_2$ ($\alpha_4$) possesses here a notably higher (reduced) frequency when 
compared to region $I$ being also larger (smaller) than the $\alpha_1$ ($\alpha_3$) [see also the discussion below].  
Inspecting the response regions $III$ and $IV$ we deduce that only the single atom tunneling mode $\alpha_1$ survives, being however significantly weakened 
due to the almost homogeneous interaction strengths ($C_{int}$) that are formed for these wavevectors. 
To examine the robustness of the above-mentioned tunneling modes in the case of a smaller interaction offset $g$, we present 
in the inset ($b_1$) $F(\omega;k_1)$ for $g=0.1$. 
As shown, due to the weak interaction offset only the single-particle tunneling mode $\alpha_1$ survives possessing 
an overall smaller frequency when compared to the $g=1$ case. 

To obtain a basic understanding on the existence of the tunneling modes induced by the quench, we 
employ a crude measure for the spatially averaged interaction energy [see Eq. (\ref{aver_int}) below] of a 
particular single-particle density distribution. 
As already mentioned in Sec. \ref{S3} different spatial configurations (symmetric around $x=0$) of the 
interaction strength with respect to $k_1$ cause   
only parity symmetric (with respect to the central well) number states to be energetically equal.  
Focussing exclusively on the lowest band interwell tunneling, we roughly approximate 
the spatially averaged interaction energy of a particular particle configuration characterized by the number state $\ket{\vec{n}}=\ket{n_1,n_2,n_3}$ as    
\begin{equation}
\begin{split}
&\bar{E}^{int}_{|\vec{n}\rangle}(g,a,k_1,\phi)=\\&\sum_{i=1}^m \; \frac{n_i(n_i-1)}{2(d_i-d'_i)^2} \int_{d_i}^{d'_{i}} dx \; C_{int}(x;g,a,k_1,\phi). \label{aver_int} 
\end{split}
\end{equation} 
$n_i$ refers to the number of bosons located at the $i$-th well, $d_i$, $d'_i$ denote 
the edges of the $i$-th well and $N$ is the total number of bosons.  
Then, the spatially averaged interaction energy difference of the different Fock states (modes) is approximately determined 
by $\Delta \bar{E}^{int} = \bar{E}_{|\vec{n}\rangle}^{int} -\bar{E}_{|\vec{n}'\rangle}^{int}$.  
For the tunneling modes participating in our system, the corresponding spatially averaged energy differences, $\Delta \bar{E}^{int}$ 
are illustrated in Fig. \ref{fig:fidelity1_a} ($c$). 
Referring to the different modes $\Delta \bar{E}^{int}(k_1)$ schematically resembles the energetic order of the frequency branches 
depicted in Fig. \ref{fig:fidelity1} ($b$) but does not provide any quantitative values as within our approximation we do not take into account the explicit form 
of the corrresponding Wannier function. 
Moreover using $\Delta \bar{E}^{int}(k_1)$ the appearence of weak and strong response regions can be explained only in some limited cases.  
Indeed, for integer values of $k_1$ $C_{int}$ is on average the same for each well, i.e. the repulsion of the atoms 
within each well is alike, and therefore $\Delta \bar{E}^{int}(k_1=n \in \mathbb{N})\to0$ for the states $|1,2,1\rangle$, $|2,1,1\rangle$ and $|1,1,2\rangle$. 
As a consequence the single-particle tunneling mode $\alpha_1$ is supressed during the dynamics. 
However, the absence of higher order tunneling processes such as $\alpha_2$-$\alpha_5$ for $k=n \in \mathbb{N}$ can not be understood utilizing $\Delta \bar{E}^{int}$ 
[see Fig. \ref{fig:fidelity1_a} ($c$)].   
On the contrary $\Delta \bar{E}^{int}$ for $k=(n+1)/2$ with $n\in \mathbb{N}$ captures qualitatively at least the behavior of the observed frequency 
spectra see e.g. the exchange of the $\alpha_1$ ($\alpha_3$) and $\alpha_2$ ($\alpha_4$) modes which is energetically favored.  
In addition, for these $k_1$ values an interaction imbalance between the middle and outer wells 
(e.g. for $k_1=0.5$ an interaction peak appears in the middle well) takes place giving rise to several tunneling modes. 

An effective way to manipulate the tunneling frequencies (within the regions $I$-$IV$) is 
to tune the spatial interaction strength by means of the inhomogeneity parameter $a$ or the interaction offset $g$. 
$F(\omega;k_1)$ for $a=5$ and $g=1$, $\phi=0$ is shown in Fig. \ref{fig:fidelity1_a} ($d$). 
Besides $\alpha_4$, all the previously observed tunneling modes ($\alpha_1$-$\alpha_3$ and $\alpha_5$) appear in the 
fidelity spectrum but are found to be shifted to larger frequencies. 
The absence of the $\alpha_4$ mode is caused by the fact that for increasing $a$ the $|1,3,0\rangle$ contribution in the ground state 
becomes negligible and therefore the corresponding tunneling process $\alpha_4$ is supressed. 
The characterization of all the above frequency branches in terms of the dominant number state transitions has been achieved via 
the multiband expansion of Eq. (\ref{multiband}). 
The observed shift of the branches $\alpha_1$-$\alpha_3$ and $\alpha_5$ can also be understood in terms of $\Delta \bar{E}^{int}(a,k_1)$ 
which acquires larger values for increasing $a$ and fixed $k_1$ as shown in Fig. \ref{fig:fidelity1_a} ($c$) for the frequency branch $\alpha_1$. 
Finally, we investigate the influence of the interaction offset $g$ on the quench induced tunneling modes, see Fig. \ref{fig:fidelity1_a} ($e$) 
for fixed $k_1=0.5$, $a=1$ and $\phi=0$. 
All five interaction dependent modes ($\alpha_1$-$\alpha_5$) occur but importantly here only the single-particle mode $\alpha_1$ survives 
for small $g<0.5$ as well as for strong $g>3$ interaction offsets.   
The latter indicates a suppression of the interwell tunneling dynamics which is highly altered in the strongly interacting regime \cite{Smerzi,Albiez}. 
Summarizing, a quench of the wavevector of the spatially-dependent interaction strength induces a multitude of tunneling modes which can 
be further amplified, diminished or shifted by adjusting individually parameters of the interaction profile. 

\subsection{Breathing dynamics} 
\label{S4_breathing} 

\begin{figure}[ht]  
  \centering
\includegraphics[width=0.45\textwidth]{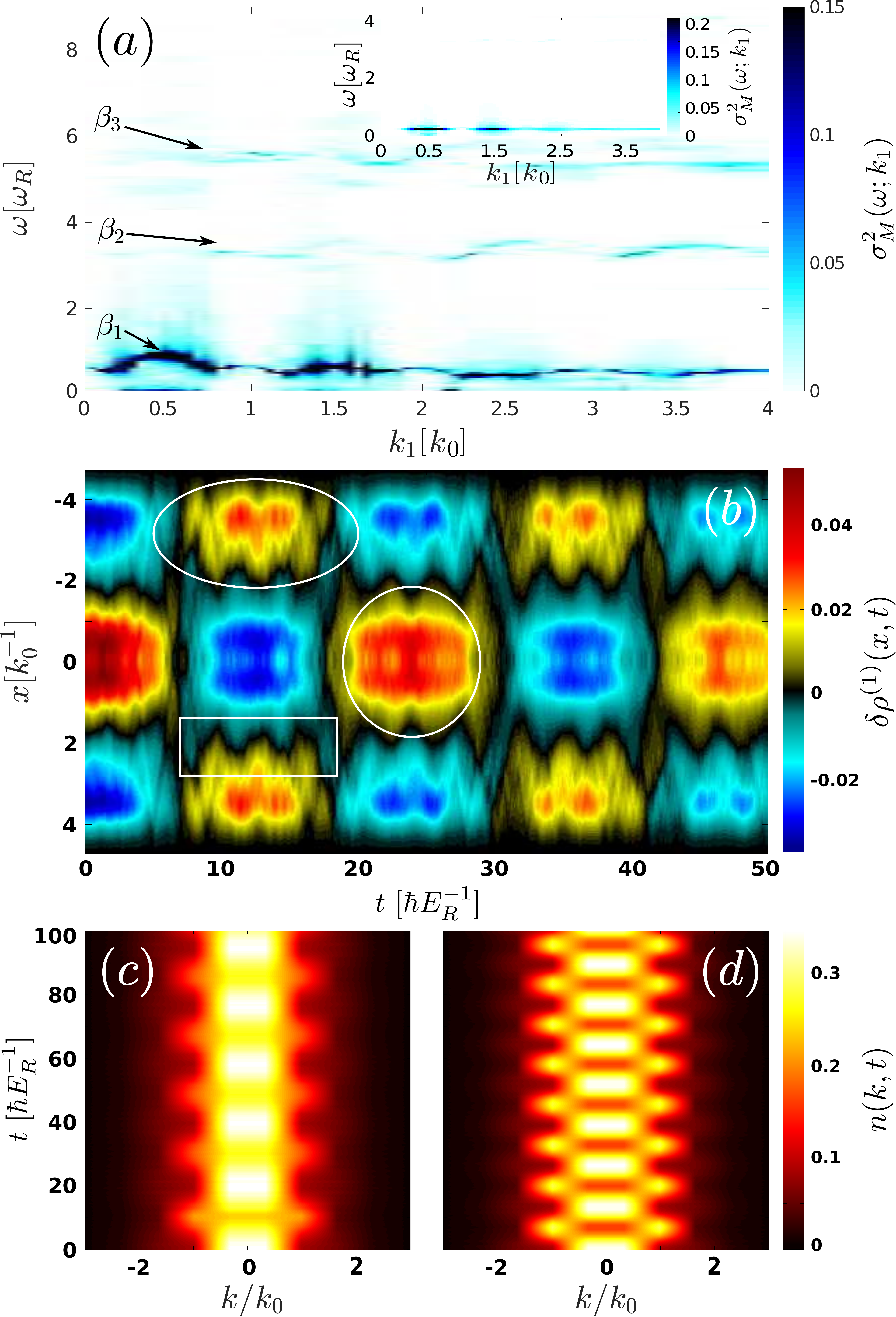}
  \caption{ ($a$) Spectrum of the variance $\sigma_M^2(\omega;k_1)$ for increasing wavevector $k_1$ of $C_{int}$ with $g=1$, $a=2$ and $\phi=0$.  
  The inset shows $\sigma_M^2(\omega;k_1)$ for the interaction offset $g=0.1$ and varying wavevector $k_1$. 
  The remaining system parameters are the same as in Fig. \ref{fig:fidelity1} ($a$).  
  ($b$) One-body density fluctutations $\delta\rho^{(1)}(x,t)$ for a wavevector quench from $k_1=0$ to $k_1=0.5$.  
  The ellipse, circle and rectangle indicate the cradle, breathing and over-barrier transport respectively. 
  ($c$), ($d$) Evolution of the momentum distribution of the one-body density matrix after a wavevector quench from $k_1=0$ to ($c$) $k_1=0.2$ and ($d$) $k_1=0.5$.  
  The horizontal axis refers to the momenta in units of the inverse lattice vector $k_0=\pi/l$. 
  In all cases the system is initialized in the ground state of four bosons confined in a triple well.}\label{fig:variance1}
\end{figure}

Having discussed in detail the lowest band tunneling dynamics trigerred by quenching the wavevector of $C_{int}$, we next investigate 
the corresponding excitation to higher band (interband) processes. 
The dominant mode here, that contains admixtures of excited band states 
corresponds to a breathing mode. 
The latter refers to an expansion and contraction of the atomic cloud \cite{Abraham_breath,Abraham_breath1} and     
due to the lattice symmetry (parity symmetry with respect to $x=0$) it is expected to be more prone 
within the central well \cite{Mistakidis,Mistakidis1}.  
To track this mode we measure the position variance in the middle well    
$\sigma^2_M(t)$ [Eq. (\ref{variance})]. 
To quantify the frequency spectrum of the breathing mode we inspect $\sigma^2_M(\omega;k_1)=\operatorname{Re} \left\{ 1/\pi\; \int dt\; \sigma^2_M(t)e^{i \omega t}\right\}$ 
with varying wavevector in Fig. \ref{fig:variance1} ($a$).  
Three distinct $k_1$-dependent frequency branches can be observed. 
To identify the dominant number state transitions that correspond to these frequency branches 
we employ the multiband expansion, see Eq. (\ref{multiband}). 
The energetically lowest branch $\beta_1$ is linked to the most dominant interwell tunneling mode 
$|1,2,1\rangle \leftrightharpoons |1,1,2\rangle$ [see also Fig. \ref{fig:fidelity1} ($b$)]. 
The appearance of this mode in the spectrum is attributed to the fact that the tunneling can 
induce a change in the width of the local wavepacket.
The branch $\beta_2$ refers to the interband process 
$|1\otimes 1^{(2)},1,1\rangle \leftrightharpoons |1,2,1\rangle \leftrightharpoons |1,1,1\otimes 1^{(2)}\rangle$ 
which indicates the occurence of a global interwell breathing mode induced by the over-barrier 
transport (i.e. the probability for a single-particle to possess enough energy to overcome the lattice barrier). 
It fluctuates around $\omega\approx3.5$ as $k_1$ is increased exhibiting maxima and minima 
within the enhanced and weak response areas respectively. 
We remark here that in order to discern between over-barrier transport and below the barrier tunneling we calculate the occupation 
probability of the number states that belong to delocalized single-particle energy bands. 
The latter bands lie above the lattice barrier and their respective Wannier states are delocalized over the lattice structure. 
Commonly e.g. within the framework of the Bose-Hubbard model \cite{Motttheo} tunneling below the barrier corresponds to transitions between well-localized energy bands, 
i.e. single-particle energy bands that reside well below the maximum of the lattice barrier. 
In this case, the corresponding Wannier states are also adequately localized within each well and non-overlaping between distinct wells. 
In the present setup the lattice depth is $V_0=6E_R$, and thus only the zeroth and the first excited bands [$\lambda=0,1$ in the expansion of Eq. (\ref{multiband})] 
can be considered to be adequately localized, i.e. they lie below the maximum of the lattice barrier. 
As a consequence the fraction of the over-barrier transport can be tracked via the probability amplitudes $\abs{\braket{\vec{n}|\Psi(t)}}^2$, with $\lambda>2$
in the multiband Wannier number state $|\vec{n}\rangle=|\otimes_{\lambda=0}^{j-1}n_1^{(\lambda)},..., \otimes^{j-1}_{\lambda=0}n_m^{(\lambda)}\rangle$. 
Other ways to identify over-barrier transport is to analyze the temporal fluctuations $\delta \rho^{(1)}(x,t)$ of the one-body density as it has been 
demonstrated in \cite{Mistakidis,Mistakidis1}, see also the discussion below, as well as by employing the two-body correlation function during the dynamics \cite{xontros3}. 
Finally, the branch $\beta_3$ indicates the participation of even energetically higher excitation processes such as 
$\ket{1,2,1} \leftrightharpoons |1,1,1\otimes 1^{(4)}\rangle$. 
It is located around $\omega\approx5.6$ showing a $k_1$-dependent behavior similar to $\beta_2$. 
The underlying process that triggers the aforementioned global breathing mode can be summarized as follows. 
Following a quench on the wavevector of $C_{int}$ the interaction imbalance between the central and outer wells becomes 
more pronounced leading to an over-barrier transport of one boson from the middle to the outer wells. 
Then, this boson performs a collision with the preexisting atom and a subsequent single excitation to the 
second energy band takes place, inducing this way the breathing mode. 

To showcase the emergent excitation modes in the spatially resolved system dynamics we present in Fig. \ref{fig:variance1} ($b$) 
$\delta \rho^{(1)}(x,t)$ [see also Eq. (\ref{fluctuations})] after quenching the wavevector of $C_{int}$ from $k_1=0$ to $k_{1}=0.5$. 
Regarding the global system's dynamics we observe the occurrence of a tunneling dynamics that corresponds to 
population transfer from the middle to the outer wells. 
Most importantly,  in the inner well dynamics two excited modes take place. 
Namely within the middle well a local breathing like mode is excited [see the circle in Fig. \ref{fig:variance1} ($b$)] manifested as a contraction 
and expansion of the bosonic cloud in the course of the evolution. 
In the outer wells the cradle mode occurs being a dipole-like oscillation of the localized wavepacket [see the ellipse in Fig. \ref{fig:variance1} ($b$)] 
which is generated by a direct over-barrier transport [see the rectangle in Fig. \ref{fig:variance1} ($b$)] as a consequence 
of the quench. 
For more details on the generation and further properties of this mode see \cite{Mistakidis,Mistakidis1,Mistakidis6,Jannis}. 
We remark that the cradle mode in this case of quenching $k_1$ is weak i.e. its amplitude is very small when compared to 
the breathing and tunneling modes. 
Therefore we do not proceed to a further analysis of this mode here (we shall revisit it in Sec. \ref{S5}). 
It is also important to note at this point that the breathing mode can be excited more efficiently 
upon quenching the wavevector of a spatially-dependent interaction strength as compared to    
a homogeneous interaction quench \cite{Mistakidis,Mistakidis1}.  
In this latter case the contribution of states higher than the second band is almost 
absent and can be triggered only for very strong interaction strength quenches i.e. $g>3.5$ \cite{Jannis}. 
However even in this case the breathing frequency is mainly insensitive to the interaction quench amplitude.   
To further study the dependence of the breathing mode on the interaction offset $g$ of $C_{int}$, we present in the inset of 
Fig. \ref{fig:variance1} $\sigma^2_M(\omega)$ 
for $g=0.1$ (the other parameters of $C_{int}$ are the same as before). 
As shown all excited states disappear and only the lowest band tunneling mode persists, since such a small offset 
is insufficient to induce over-barrier transport. 

Finally, we study the effect in momentum space of quenching the wavevector of $C_{int}$.  
To this end, we employ the momentum distribution of the one-body density $n(k,t)$ [see Eq. (\ref{momentum})], aiming 
to reveal whether certain momenta can be populated in the course of the dynamics. 
Figures \ref{fig:variance1} ($c$) and ($d$) show $n(k,t)$ following a $k_1$-quench of $C_{int}$ 
from $k_1=0$ to $k_1=0.2$ and $k_1=0.5$ respectively. 
As it can be easily deduced, $n(k,t)$ exhibits a breathing dynamics where different momenta are populated. 
In particular, $n(k,t)$ corresponds to a Gaussian like distribution centered around $k_0=0$ possessing edges either 
at $\pm k_0/2=1.57$ or $\pm 3k_0/2=4.71$ where in both cases all momenta in between are activated. 
The oscillation frequency between the above-mentioned momenta depends strongly on the quench amplitude, namely it is 
larger for $k_1=0.5$ compared to $k_1=0.2$. 
Therefore, it follows the system's dynamical response [see also Fig. \ref{fig:fidelity1} ($a$)] and in particular the tunneling dynamics. 
Similar periodically modulated patterns in the momentum distribution during the evolution take place   
when considering multiple interaction quench sequences in few boson homogeneously interacting ensembles trapped 
in an optical lattice \cite{Jannis}.

\section{Quench of the Phase of the Interaction Strength} 
\label{S5} 

Let us now examine the dynamics upon a sudden change of the phase $\phi$ of the spatially-dependent interaction strength $C_{int}$. 
Following this quench protocol an interaction imbalance between all wells of the lattice is induced resulting 
in a directed tunneling dynamics (see also below). 
The system (four bosons in a triple well) is initially prepared in the ground state of the Hamiltonian of Eq. (\ref{Hamiltonian}) 
with $C_{int}$ of Eq. (\ref{Eq:interactioncoefficient}) characterized by $g=1$, $k_1=0.2$, $a=2$ and $\phi=0$. 
Namely at $t=0$, $C_{int}= 1+2\;\cos^2(0.2x)$. 
At $t=0$ a quench on the phase of $C_{int}$ is performed and the Hamiltonian that governs the dynamics is again described by Eq. (\ref{Hamiltonian}) but 
the spatially-dependent effective interaction is $C_{int}= 1+2\;\cos^2(0.2x+\phi)$.
Then the spatially averaged in each well $C_{int}$ possesses a maximum value around the central well 
and resembles an almost linear gradient. 
We remark here that for larger $k_1$ values being comparable to the lattice wavevector $k_0$ 
a phase quench does not produce a substantial dynamical response.  
The dominant contribution in the initial many-body state stems from the number state $|1,2,1\rangle$. 

\subsection{Tunneling dynamics} 
\label{S5A} 

\begin{figure}[ht]  
  \centering
\includegraphics[width=0.5\textwidth]{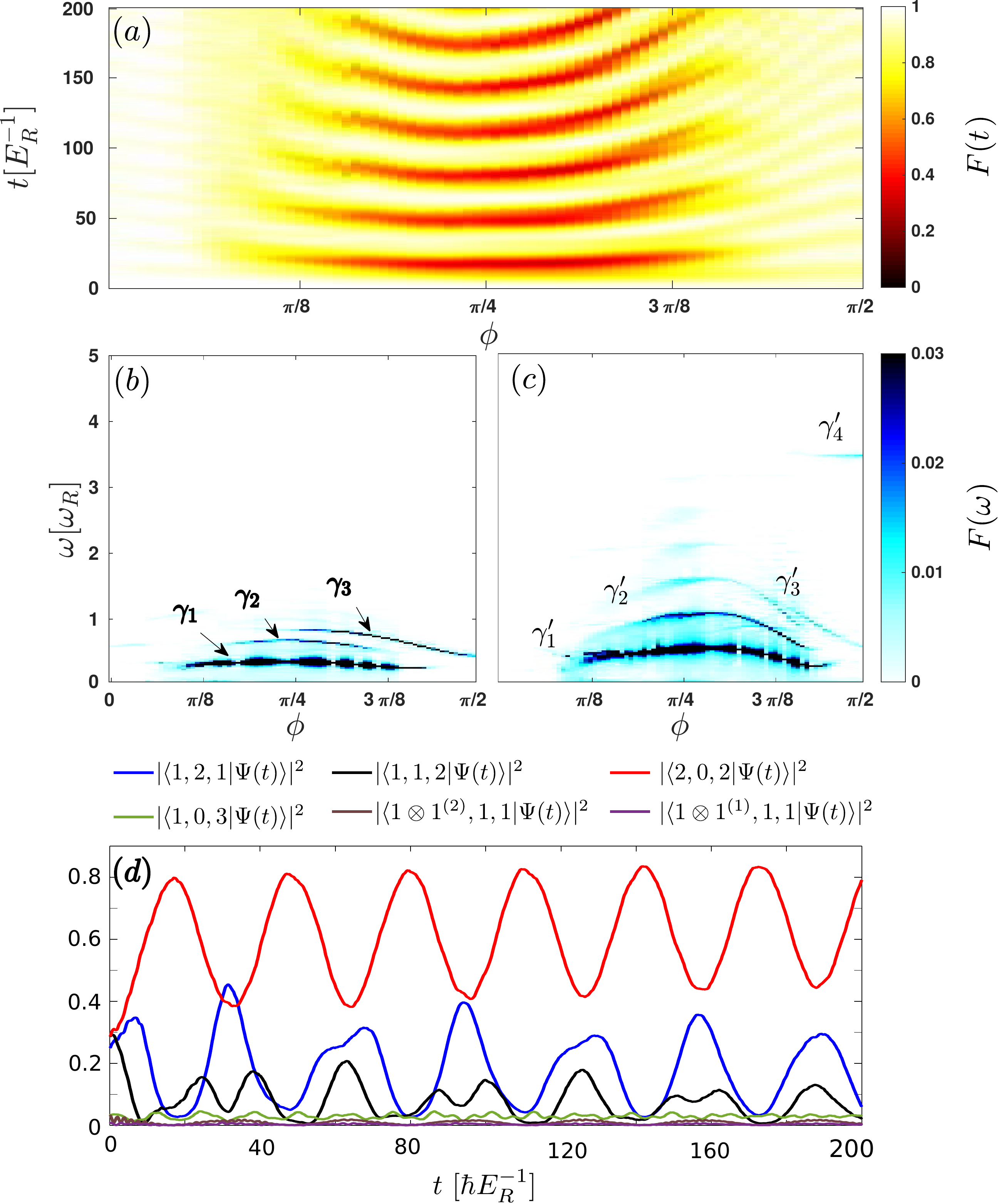}
  \caption{ ($a$) Fidelity evolution $F(t;\phi)$ for a varying phase $\phi$ of the interaction strength $C_{int}$.  
  The corresponding spectra $F(\omega;\phi)$ for inhomogeneity ($b$) $a=2$ and ($c$) $a=5$. 
  ($d$) Time-evolution of different number state configurations (see legend) upon quenching the phase from $\phi=0$ to $\phi=\pi/4$.  
  The other parameters of the spatially-dependent interaction strength $C_{int}$ are $g=1$, $a=2$ and $k_1=0.2$. 
  In all cases the system is initially prepared in the ground state of four bosons in a triple well.} 
  \label{fig:fidelity2}
\end{figure}

To examine the response of the system induced by a quench of the phase $\phi$ of the interaction strength $C_{int}$ we 
invoke the fidelity evolution $F(t;\phi)=|\langle \Psi(0)|\Psi(t;\phi) \rangle|^2$, see Fig. \ref{fig:fidelity2} ($a$). 
Due to the underlying periodicity of $C_{int}$ [see also Eq. (\ref{Eq:interactioncoefficient})] we restrict our study 
to the phase interval $[0,\pi/2]$.  
Recall here that for $\phi=0$ ($\phi=\pi/4$) the spatially averaged in each well interaction strength 
exhibits a maximum around the central (left) well, while 
at $\phi=\pi/2$ we encounter a corresponding minimum in the middle well [see also Fig. \ref{fig:interactionconfiguration}].   
The system is significantly perturbed i.e. $F(t;\phi)\neq1$ (unperturbed, $F(t;\phi)\approx1$) for $\pi/16<\phi<\pi/2$ ($\phi<\pi/16$) 
where the quench induced spatial interaction imbalance becomes significant (negligible). 
The underlying interaction imbalance is strongest within the range $\pi/8 \leq \phi \leq 3\pi/8$ (maximized at $\phi=\pi/4$) and as a consequence the system is strongly driven 
out-of-equilibrium. 
$F(t;\phi)$ exhibits oscillations which possess the largest amplitude in the vicinity of $\phi \approx \pi/4$. 

To identify the participating modes triggered by the phase quench in $C_{int}$ we inspect the fidelity spectrum $F(\omega;\phi)$ shown 
in Fig. \ref{fig:fidelity2} ($b$). 
Three distinct tunneling pathways occur in the spectrum, denoted by $\gamma_1$, $\gamma_2$ and $\gamma_3$, which can be linked 
to first and second order transport. 
In particular, the two lowest-lying phase-dependent frequency branches refer to the first order processes 
$\ket{1,2,1} \leftrightharpoons \ket{1,1,2}$ ($\gamma_1$) and $\ket{1,2,1} \leftrightharpoons \ket{2,1,1}$ ($\gamma_2$) respectively. 
The assignment of these transitions referring to the dominantly contributing number states, in the course of the dynamics, has been achieved 
by utilizing the multiband number state basis of Eq. (\ref{multiband}). 
As an illustration we show in Fig. \ref{fig:fidelity2} ($d$) the probabilities $\abs{\braket{\vec{n}|\Psi(t)}}^2$ for the predominantly 
populated number states when considering a phase quench from $\phi=0$ to $\phi=\pi/4$. 
In order to identify the transitions between number states that correspond to the frequency branches appearing in $F(\omega,\phi)$ we calculate the 
spectrum of $\abs{\braket{\vec{n}|\Psi(t)}}^2$ for each number state transition and then match its frequency to the branch occuring in $F(\omega,\phi)$. 
Of course, this procedure has been followed for all quench amplitudes within the range $\phi\in \{0,\pi/2 \}$ (not shown here).

The above-mentioned tunneling processes that would otherwise (for $C_{int}$ with $\phi=0$) be energetically equal are well separated here as a consequence of the employed quench protocol. 
Indeed quenching the phase of $C_{int}$ the prevailing interatomic interaction is shifted from the middle to the left well. 
Therefore the atoms initially located at the left site experience increasing repulsion after the quench rendering 
the single-particle tunneling from the middle to the right well energetically favorable (branch $\gamma_1$).   
In this way a directed tunneling process can be achieved.  
Another interesting observation here is that the frequencies of the tunneling modes $\gamma_1$ and $\gamma_2$ 
start to merge into a single one for $3\pi/8<\phi<\pi/2$ as the corresponding spatially averaged interaction strength in the left and right 
wells becomes comparable within this phase interval.  
Finally, we encounter the second order tunneling process $|1,2,1\rangle \leftrightharpoons |2,0,2\rangle$ (indicated by the branch $\gamma_3$)  
which is more pronounced for $\pi/4<\phi<\pi/2$ where the interaction imbalance is most pronounced when compared to the $0<\phi<\pi/4$ phase interval.   

In an attempt to steer the above-mentioned tunneling modes or even trigger higher-lying ones we present in Fig. \ref{fig:fidelity2} ($c$) 
$F(\omega;\phi)$ for a larger inhomogeneity amplitude $a=5$. 
The resulting lowest band tunneling modes ($\gamma'_1$, $\gamma'_2$ and $\gamma'_3$) are the same as before but they 
are located at higher frequencies, they become stronger and the frequency gap between each two is more pronounced when compared to the case of $a=2$.  
In addition we observe the existence of the higher lying frequency branch $\gamma'_4$ which refers to the interband transition 
$\ket{1,2,1}\leftrightharpoons\ket{1,1,1\otimes 1^{(2)}}$.  
This mode located in the vicinity of $\phi=\pi/2$ is caused by the enhanced over-barrier transport occuring for these 
quench amplitudes. 
Note that such an interband transition is also inherently related to the considered large interaction inhomogeneity being, in general, supressed 
for smaller inhomogeneities, see for instance Fig. \ref{fig:fidelity2} ($b$). 
In the following subsection, we explicitly address how to excite such higher band states for varying inhomogeneity $a$.  

\subsection{Excitation processes} 
\label{S5B}

Having discussed in detail the tunneling mechanisms upon a phase quench, we next investigate the possibility of triggering interband transitions. 
As already mentioned above the phase quench shifts the interaction profile $C_{int}$ in space. 
Here $C_{int}$ initially ($\phi=0$) exhibits a maximum in the central well and after quenching $\phi$ this maximum moves to the left well thus inducing 
the aforementioned tunneling dynamics. 
In addition the sudden change of $\phi$ yields a high probability for the delocalized particle to overcome 
the lattice barrier (over-barrier transport) and move to the neighboring well in which the minimum of the postquench $C_{int}$ occurs. 
The resulting over-barrier transport is consequently responsible for mainly two higher band excited modes, 
namely the global breathing \cite{Mistakidis,Koutentakis,Tschischik} and the local cradle \cite{Mistakidis,Mistakidis1} modes. 

\begin{figure}[ht]  
  \centering
\includegraphics[width=0.5\textwidth]{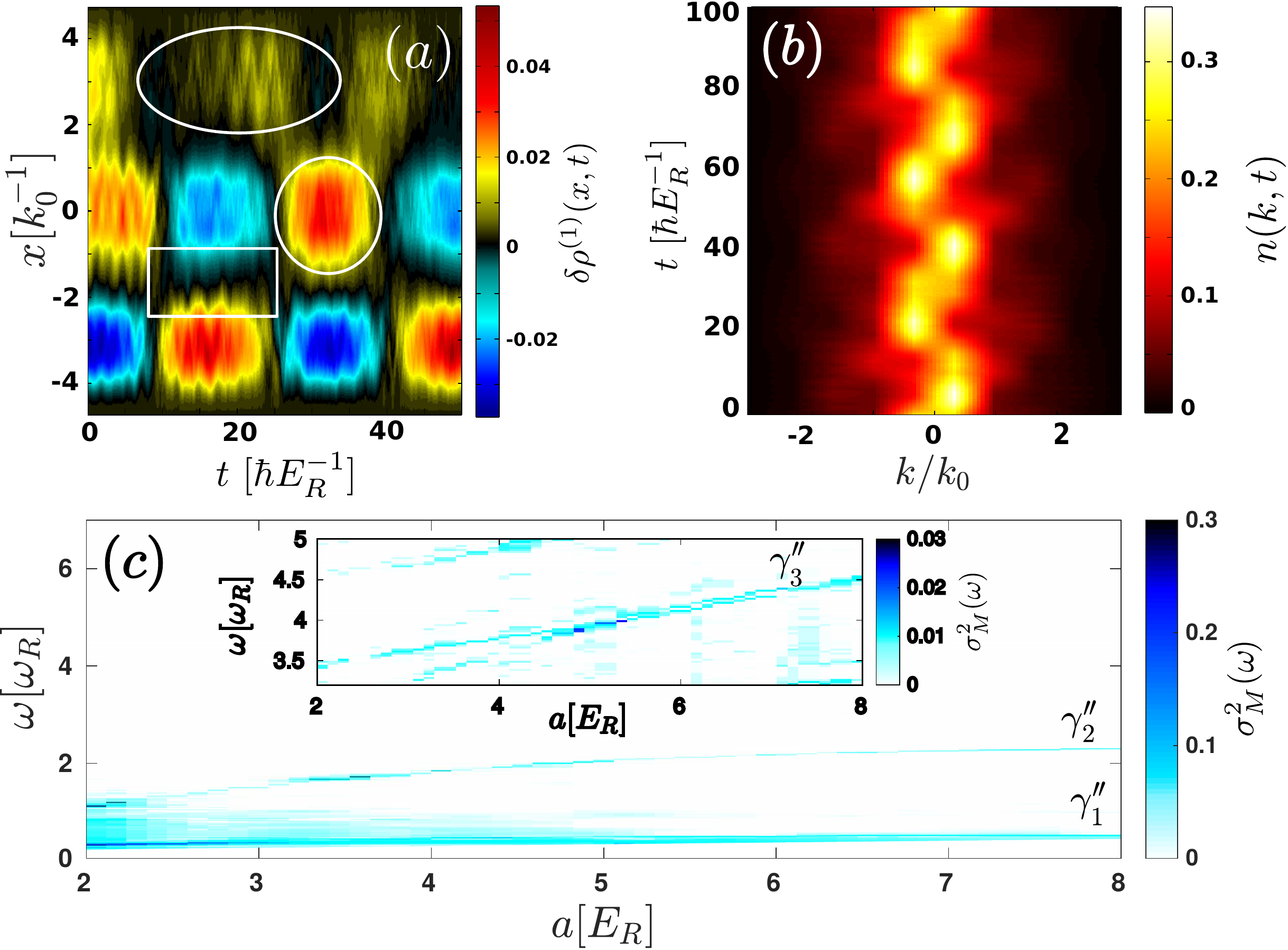}
  \caption{ ($a$) $\delta\rho^{(1)}(x,t)$ and ($b$) $n(k,t)$ following a phase quench from $\phi=0$ to $\phi=\pi/4$.   
  Ellipse, circle and rectangle mark the cradle, breathing and over-barrier transport respectively.   
  The horizontal axis in ($b$) represents the momenta in units of the inverse lattice vector $k_0=\pi/l$. 
  The other parameters of $C_{int}$ correspond to $g=1$, $a=2$ and $k_1=0.2$.  
  ($c$) Spectrum of the intrawell asymmetry $\Delta\rho_R(\omega)$ and the variance $\sigma_M^2(\omega)$ [inset ($d_1$)] for increasing 
  inhomogeneity $a$. 
  The spatially-dependent interaction profile $C_{int}$ is characterized by $g=1$, $\phi=\pi/8$ and $k_1=0.2$.
  The system is initially prepared in the ground state of four bosons in a triple well.}  
  \label{fig:fidelity2_a}
\end{figure}

All the aforementioned modes can be visualized in the dynamics of $\delta \rho^{(1)}(x,t)$, 
shown in Fig. \ref{fig:fidelity2_a} ($a$) after a phase quench from $\phi=$ to $\phi=\pi/4$. 
We observe that predominantly a population transfer (tunneling mode) from the middle well to the right half of the triple well, located at $0<x<4.8$, 
and back occurs. 
Additionally, the intrawell dynamics is mainly comprised by two excited modes. 
A breathing mode in the middle well and a cradle mode in the outer ones take place with both being induced by the 
over-barrier transport caused by the quench, see in particular the circle, ellipse and rectangle 
in Fig. \ref{fig:fidelity2_a} ($a$) respectively. 

To gain further insight we inspect the dynamics in momentum space by 
employing the one-body density momentum distribution $n(k,t)$. 
Figure \ref{fig:fidelity2_a} ($b$) presents $n(k,t)$ following a phase quench of $C_{int}$ from $\phi=0$ to $\phi=\pi/4$. 
As shown, a sudden change of the phase of $C_{int}$, yields a directed tunneling motion to the right side of the triple well ($0<x<4.8$), 
see also Fig. \ref{fig:fidelity2_a} ($a$). 
More specifically, we observe first a consecutive population of $k_0=0$, $k_0/2=1.57$ and $k_0=3.14$ momenta, e.g. see $n(k,t=10)$, 
and subsequently of the exactly opposite ones, see $n(k,t=20)$. 
This process is repeated during the evolution in a periodic manner. 
Finally, we remark that different quench amplitudes affect mainly the speed of the alternating 
activation of momenta and to a lesser extent the magnitude of their population (not shown here for brevity reasons), 
see also our discussion in Appendix A for the five well case and in particular Figs. \ref{fig:largersystem} ($o$) and ($m$). 

It is important to mention here that the cradle mode occurs also upon a wavevector quench but it is greatly supressed when 
compared to the phase quench scenario.  
In this latter case the interaction imbalance between individual wells is stronger resulting in an enhanced over-barrier transport and  
thus a prone cradle process. 
The cradle mode represents a dipole-like intrawell oscillation in the outer wells of the finite 
lattice (for more details on the generation and properties of this mode see \cite{Mistakidis,Mistakidis1}).  
Since the parity symmetry within the outer well where it takes place is broken, it can be quantified by the 
corresponding intrawell asymmetry of the wavefunction. 
Since here we are interested in the right well dynamics the aforementioned asymmetry is defined as 
$\Delta\rho_R(t)=\rho_{R,1}(t)-\rho_{R,2}(t)$, where $\rho_{R,1}(t)$ and $\rho_{R,2}(t)$ 
denote the spatially integrated densities of the left and the right half sector of the well. 
To trace this mode we rely on $\Delta\rho_R(\omega)$. 
Finally, the global breathing mode refers to the contraction and expansion of the entire bosonic cloud being induced by the 
over-barrier transport. 
Due to the lattice symmetry [see also Sec. \ref{S4_breathing}] the global breathing mode is expected 
to be more prone in the central well and therefore $\sigma_M^2(\omega)$ provides an adequate measure for this mode. 

As both of the above described modes are initialized by the over-barrier transport which in turn depends 
on the inhomogeneity $a$ it would be instructive to study how they are affected by adjusting $a$.  
Let us therefore inspect them by focussing on a specific phase quench from $\phi=0$ to $\phi=\pi/8$ and considering fixed $k_1=0.2$ and $g=1$ for 
varying $a$. 
Fig. \ref{fig:fidelity2_a} ($c$) presents $\Delta\rho_R(\omega;a)$ and $\sigma_M^2(\omega;a)$ [see the inset of Fig. \ref{fig:fidelity2_a} ($c$)] for increasing $a$.   
Regarding the cradle mode (see branch $\gamma_2''$), it can be linked to the interband transition 
$|1,2,1\rangle \leftrightharpoons |1,1,1\otimes1^{(1)}\rangle$ and it is greatly affected by the 
considered inhomogeneity amplitude $a$. 
This latter behavior is expected as an increasing $a$ triggers all the more the over-barrier transport. 
In particular, its characteristic frequency (branch $\gamma_2''$) increases for larger inhomogeneity amplitudes $a$. 
The observed energetically lowest branch $\gamma_1''$ corresponds to the tunneling process $|1,2,1\rangle \leftrightharpoons |1,1,2\rangle$ which 
is only weakly $a$-dependent. 
On the other hand, the global breathing mode [see branch $\gamma_3''$ in the $\sigma_M^2(\omega;a)$] refers to the interband tunneling 
$|1,2,1\rangle \leftrightharpoons |1,1,1\otimes 1^{(2)}\rangle$ and it increases almost linearly for varying $a$ due to the 
consequent enhanced interaction imbalance between the adjacent wells.  
Summarizing, by tuning the inhomogeneity of the spatially-dependent interaction strength we can manipulate the 
frequencies of both excited higher band modes. 

To generalize our findings, in Appendix A, we demonstrate that the main characteristics of the dynamical response upon quenching either the 
wavevector or the phase of the spatially-dependent interaction strength remain robust also in the case 
of six bosons trapped in a five well lattice i.e. with filling $\nu>1$. 
Let us mention that we have checked that a similar dynamical response occurs also for larger systems e.g. seven bosons in six wells (results not shown here). 
It is important to remark here that for fillings $\nu<1$ a corresponding quench of the spatial interaction profile does not 
alter significantly the initial (ground) state of the system (results not shown due to brevity) as the overlap between the individual 
bosons is small.

\section{Conclusions} 
\label{conclusion}

We have investigated the ground state properties and in particular the 
nonequilibrium quantum dynamics of few boson ensembles experiencing a spatially 
modulated interaction strength and confined in a finite lattice potential. 
To profit from the competition between delocalization and on-site interaction effects we focus on setups  
possessing fillings larger than unity, thus also avoiding suppression of tunneling.  
The employed spatial interaction strength is of sinusoidal type and it is characterized by its modulation wavevector,  
inhomogeneity amplitude, interaction offset and phase.   

Before delving into the dynamics, we trace the impact of the wavevector and the phase individually on the 
ground state properties of the system. 
The inhomogeneity amplitude in most cases is kept fixed being of the order of half the lattice depth, while the interaction offset is unity. 
For small values of the wavevector the spatially in each well averaged interaction strength is larger 
within the central well when compared to the outer ones, while it becomes the same for incrementing spatial periodicity.   
This behavior causes a spatial redistribution of the atoms from the outer to the central wells for increasing wavevectors.  
In all cases, the ensemble remains superfluid. 
On the other hand, phase shifts yield an interaction imbalance between all lattice wells  
and enables us to displace the single-particle density distribution in a preferred direction achieving Mott-like states. 

Next, we analyze the system's dynamical response upon quenching either the wavevector or the phase of the spatial interaction strength. 
Following a sudden change of the wavevector the dynamics is characterized by enhanced response regions, located at fractional values of the wavevector,  
in which bosons at distinct wells are subjected to different spatially averaged interaction strengths.   
For incrementing wavenumbers these enhanced response regions become gradually less transparent as the respective interaction profile 
tends to a homogeneous configuration. 
The quench on the wavevector of the spatially-dependent interaction strength yields the excitation of a multitude 
of tunneling modes consisting of single and two particle transport.   
These modes can be further amplified or shifted by adjusting the interaction offset or the inhomogeneity amplitude respectively. 
A quench induced breathing dynamics is also observed characterized by interband tunneling processes which possess mainly a single excitation 
to the second or fourth excited band. 
We also note that a cradle interband excitation mode occurs which, however, possesses a very weak amplitude when compared to the breathing mode thus rendering the latter 
the dominant higher band excitation process. 
Inspecting the momentum distribution we show that a periodic population transfer of momenta during the dynamics takes place, while 
the one-body coherence function reveals that partially coherent regions occur between the wells that are predominantly populated during the evolution. 

The phase quench imposes an interaction strength imbalance between all wells yielding a directed transport along the finite lattice as it 
accounts for a spatial shift of the entire interaction profile.  
The induced transport consists of single-particle and atom pair tunneling. 
More importantly and in contrast to the wavevector quench, a phase quench allows for the discrimination of the 
tunneling modes which would be otherwise energetically equal.  
A characteristic process of the latter type corresponds for instance to single-particle lowest band tunneling from the middle 
to the left or the right well. 
For larger inhomogeneity amplitudes these modes become more discernible as a function of the phase parameter, namely their energy 
difference is intensified, while for increasing interaction offset their supression is observed. 
The directional transport is also reflected in the evolution of the one-body momentum distribution and the coherence function. 
In the former case a directed consecutive population of higher momenta occurs, while in the latter case the predominantly 
populated wells form a partially incoherent region which is shifted in the preferred tunneling direction. 
Besides the above described tunneling dynamics, the phase quench yields a noticeable over-barrier transport which in turn 
induces a global breathing motion of the entire bosonic cloud and a cradle mode in the outer wells. 
Both modes are related to single-particle interband processes to the first or second excited band respectively, and are 
found to be enhanced for incrementing inhomogeneity amplitude. 

There are several directions that one might pursue as possible extensions of the present work. 
An intriguing prospect would be to study the periodically driven dynamics upon shaking the optical 
lattice and investigate how the properties of the corresponding parametrically induced resonances are altered when compared to  
the homogeneously interacting case \cite{Mistakidis4}.  
Another possible path is to explore the nonequilibrium dynamics of bosonic binary mixtures experiencing such spatially 
dependent interactions. 
Here it is interesting to unravel whether a phase separation process can be achieved after quenching the wavevector  
of the interaction profile and even analyze the triggered excitation modes.

\appendix

\section*{Appendix A: Quench dynamics in a five well optical lattice} 
\label{five_wells}

Let us consider six bosons confined in a five well finite optical lattice. 
The system is initially prepared in its ground state where the corresponding spatial interaction strength 
[see also Eq. (\ref{Eq:interactioncoefficient})] is characterized by $g=1$, $\phi=0$, $a=2$, $k_1=0$  
[$a=3$, $k_1=0.05$] for the wavevector [phase] quench.    
Then, the initial state is an admixture of the available lowest band states from which the main contribution 
stems from the Wannier number states $|1,1,2,1,1\rangle$, $|1,2,1,1,1\rangle$ and $|1,1,1,2,1\rangle$.   

\begin{figure*}[ht]  
  \centering
\includegraphics[width=1.0\textwidth]{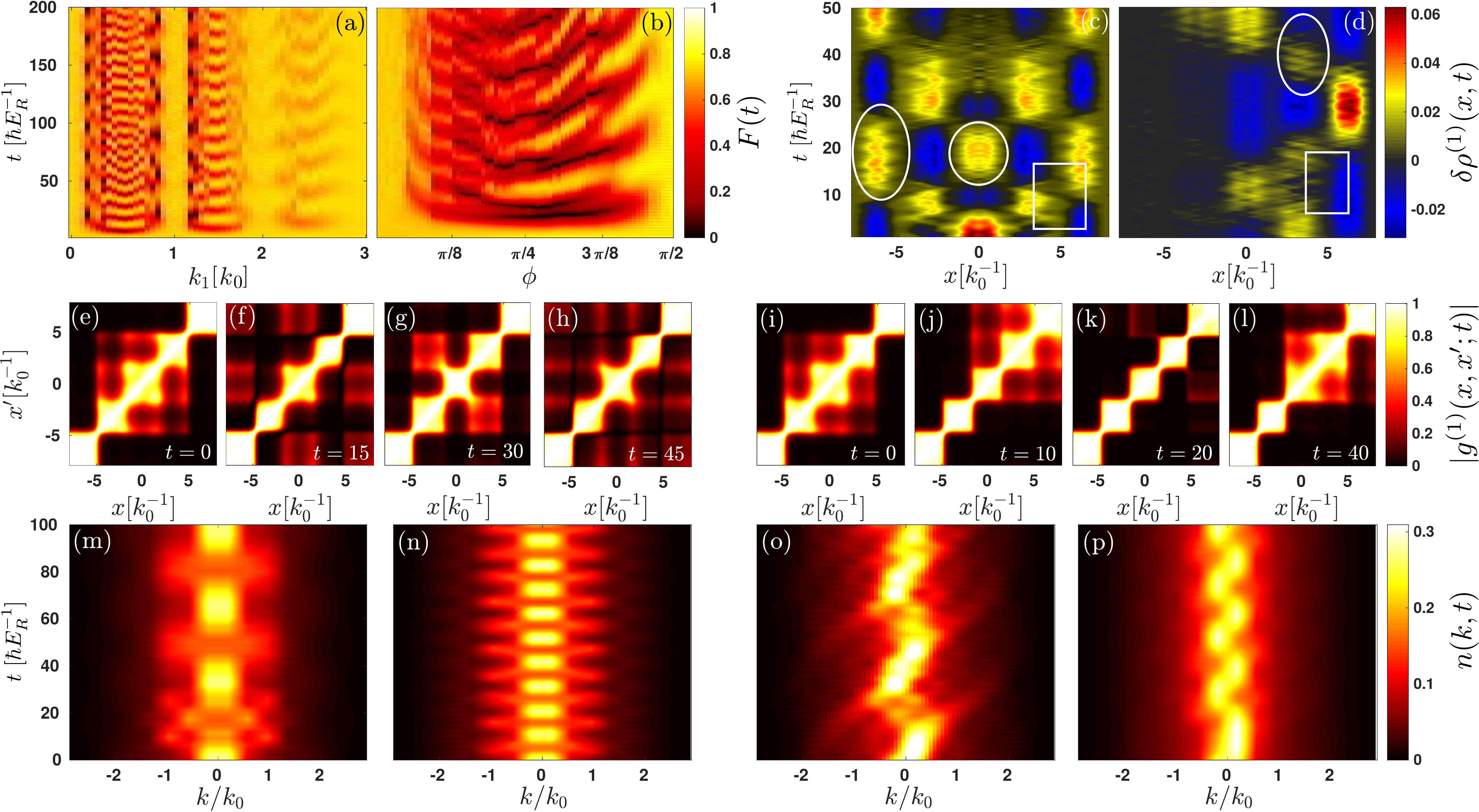}
  \caption{Fidelity evolution following a quench of ($a$) the wavevector $k_1$ and ($b$) the phase $\phi$ of the spatial 
  interaction strength $C_{int}$.   
  One-body density fluctutations $\delta\rho^{(1)}(x,t)$ for ($c$) a wavevector quench from $k_1=0$ to $k_1=0.75$ and ($d$) a phase quench from 
  $\phi=0$ to $\phi=\pi/4$. 
  One-body coherence function for distinct time instants (see legend) ($e$)-($f$) [($i$)-($l$)] after a sudden change of the magnitude of the wavevector 
  [phase] as in ($c$) [($d$)]. 
  ($m$)-($p$) Momentum distribution of the one-body density matrix during the evolution.  
  ($m$), ($n$) [($o$), ($p$)] correspond to quenches from $k_1=0$ [$\phi=0$] to $k_1=0.75$ and $k_1=0.5$ [$\phi=\pi/4$ and $\phi=\pi/8$] respectively. 
  The horizontal axis represents the momenta in units of the inverse lattice vector $k_0=\pi/l$. 
  For all cases referring to wavevector [phase] quenches the remaining system parameters correspond to $g=1$, $a=2$ and $\phi=0$ [$a=3$, $k_1=0.05$].  
  The setup consists of six bosons confined in a five well lattice. 
  The ellipses, circles and rectangles in ($c$), ($d$) indicate the cradle, breathing and over-barrier transport respectively. } 
  \label{fig:largersystem}
\end{figure*}

To induce the dynamics we perform a quench either on the wavevector $k_1$ or the phase $\phi$ of the initial interaction strength. 
Fig. \ref{fig:largersystem} ($a$) presents $F(t;k_1)$ following a wavevector quench.  
As in the triple well case, the dynamics exhibits enhanced [weak] response regions namely $F(t;k_1)\ll1$ [$F(t;k_1)\approx1$] in the 
neighborhood of $k_1=d/2$ with $d=1,3,...$ [$k_1=n \in \mathbb{N}$] due to the large [small] quench induced interaction imbalance 
of bosons residing in the different wells. 
However, the enhanced response areas appear to be wider with respect to $k_1$ when compared to the triple well case. 
Moreover, in these strong response regions $F(t;k_1)$ undergoes oscillations in time possessing a multitude of frequencies and large amplitudes  
for small $k_1$'s which tend to a single frequency oscillation of small amplitude for increasing $k_1$.   
This latter behavior is again (as in the case of the triple well) caused by the tendency of $C_{int}$ to an almost spatially 
homogeneous interaction strength on average for large $k_1$'s.    
Focussing on a phase quench, see Fig. \ref{fig:largersystem} ($b$), the obtained response resembles the triple well case 
[compare Fig. \ref{fig:largersystem} ($b$) with Fig. \ref{fig:fidelity2} ($a$)].  
Indeed, the system is driven far away from its initial state i.e. $F(t;\phi)\neq1$ exhibiting an oscillatory behavior 
for $\pi/16<\phi<\pi/2$ where the quench induced spatial interaction imbalance between distinct lattice regions becomes significant.  
This interaction imbalance is maximum at $\phi=\pi/4$ for which the oscillations of $F(t;\phi)$ possess the largest amplitude. 

In both quench scenarios, tunneling and over-barrier transport between the distinct wells of the finite lattice can be observed. 
To visualize the spatially resolved system dynamics we invoke 
$\delta \rho^{(1)}(x,t)$ \cite{Mistakidis,Mistakidis1}, see also Eq. (\ref{fluctuations}). 
Figs. \ref{fig:largersystem} ($c$), ($d$) present $\delta \rho^{(1)}(x,t)$ following a quench of the 
wavevector, from $k_1=0$ to $k_1=0.75$, and the phase, from $\phi=0$ to $\phi=\pi/4$, respectively.   
Regarding the wavevector quench, see Fig. \ref{fig:largersystem} ($c$), we observe that predominantly 
a tunneling dynamics takes place which refers to the transfer of population from the middle to the outer wells.  
Moreover, the inner well dynamics is mainly described by two excited modes. 
Specifically, the middle well exhibits a breathing like mode due to the lattice symmetry, while in the outer wells the cradle mode is  
manifested as a dipole-like oscillation of the localized wavepacket which is generated by a direct over-barrier transport as a consequence 
of the quench. 
Turning to the phase quench scenario, illustrated in Fig. \ref{fig:largersystem} ($d$), a directed population transfer from the middle well to the 
right side of the lattice located at $0<x<7.8$ (tunneling mode) and back occurs. 
Additionally, the induction of over-barrier transport caused by the quench gives rise to the cradle 
mode within the right side wells. 

To provide a link between the quench induced tunneling dynamics and the correlation properties of the system we study 
$|g^{(1)}(x,x';t)|$ at distinct time instants during the evolution \cite{Naraschewski,Sakmann_cor}.  
As already mentioned in Sec. \ref{S3}, $|g^{(1)}(x,x';t)|$ is bound to the range $[0,1]$ and 
measures the proximity of the many-body state to a product mean-field state for a fixed set of coordinates $x$, $x'$. 
Figs. \ref{fig:largersystem} ($e$)-($h$) present $|g^{(1)}(x,x';t)|$ for distinct time instants 
following a wavevector quench from $k_1=0$ to $k_1=0.75$. 
Initially, $t=0$, all bosons reside in the three central wells [see also Fig. \ref{fig:largersystem} ($c$)] which 
are partially incoherent with each other, e.g. $|g^{(1)}(x=2,x'=-4;t)|\approx0.6$ 
as depicted in Fig. \ref{fig:largersystem} ($e$).  
During evolution, an atomic portion gradually tunnels to the edge wells with the remaining atoms residing  
in the central well resulting in a low population of its nearest neighbors ($-4.7<x<-1.57$ and $1.57<x<4.7$) ones, 
see for instance Fig. \ref{fig:largersystem} ($c$) at $t=15$. 
These most outer populated wells appear to be partly incoherent [Fig. \ref{fig:largersystem} ($f$)] with each 
other (e.g. $|g^{(1)}(x=6,x'=-6;t=15)|\approx0.3$) as well as with the central well ($|g^{(1)}(x=0,x'=-6;t=15)|\approx0.5$).    
A revival of the tunneling process with population transfer from the central to the proximal to it outer wells occurs at later evolution 
times, e.g. at $t=30$ [Fig. \ref{fig:largersystem} ($g$)]. 
In turn, a partial coherence between these wells [e.g. $|g^{(1)}(x=4,x'=-4;t=30)|\approx0.6$ in Fig. \ref{fig:largersystem} ($g$)] is observed.    
Then the atoms move again to the most outer wells, e.g. at $t=45$, where the system's coherence properties [Fig. \ref{fig:largersystem} ($h$)]  
are similar to $t=15$.  
Next, we focus on the coherence properties upon quenching the phase of $C_{int}$, shown in Figs. \ref{fig:largersystem} ($i$)-($l$). 
The initial ($t=0$) partially incoherent region consists of the three middle wells [Fig. \ref{fig:largersystem} ($i$)]. 
Then, it shifts across the diagonal of $|g^{(1)}(x,x';t)|$ [at $t=10$, see Fig. \ref{fig:largersystem} ($j$)] 
including the two outer right ($1.57<x<7.8$) and the central well. 
Finally, it turns back [at $t=40$, see Fig. \ref{fig:largersystem} ($l$)] 
occupying the middle and its nearest neighbor ($1.57<x<4.7$) right well.  
As before, this behavior resembles the corresponding tunneling dynamics, see Fig. \ref{fig:largersystem} ($d$). 
Concluding from the above, we can infer that quenching the spatial interaction strength it is possible to induce either 
site selective partial coherence or even completely shift certain partially incoherent regions following the tunneling dynamics. 
Such a site selective coherence has been recently demonstrated for the ground state of a many-body bosonic ensemble trapped 
in a tilted triple well \cite{Dutta_triple}.  

Finally, we inspect whether a certain multitude of momenta is populated during the dynamics 
as a consequence of the employed quench protocol.  
To achieve the latter we rely on $n(k,t)$ \cite{Bloch,Bucker,Will}. 
The time evolution of the momentum distribution for six bosons confined in a five well 
lattice potential after a quench of the wavevector of $C_{int}$ 
from $k_1=0$ to $k_1=0.75$ and $k_1=0.5$ is depicted in Figs. \ref{fig:largersystem} ($m$), ($n$) respectively.  
As it can be seen, under this quench protocol $n(k,t)$ features in time a periodically modulated pattern 
in which distinct momenta are populated.  
In particular, $n(k,t)$ forms a gradually transformed in time broad Gaussian like distribution centered around $k_0=0$ with edges either 
at $\pm k_0/2=1.57$ or $\pm 3k_0/2=4.71$ where in both cases all momenta in between are activated. 
The oscillation frequency between the above-mentioned momentum structures changes with respect to the quench amplitude, e.g. it is 
larger at $k_1=0.5$ than $k_1=0.75$, reflecting this way the system's dynamical response [see also Fig. \ref{fig:largersystem} ($a$)] 
and more specifically the tunneling dynamics. 
We remark here that similar periodically modulated patterns in the momentum distribution during the evolution take place   
when considering multiple interaction quench sequences in few boson homogeneously interacting ensembles trapped in an optical lattice \cite{Jannis}.  
Following a sudden change of the phase of $C_{int}$, see Figs. \ref{fig:largersystem} ($o$), ($m$) for a 
quench from $\phi=0$ to $\phi=\pi/4$ and $\phi=\pi/8$ respectively, $n(k,t)$ shows a completely different behavior.  
Due to the phase shift of the interaction strength a tendency for directed tunneling to the right side of the lattice ($0<x<7.8$) occurs, 
see also Fig. \ref{fig:largersystem} ($d$). 
The latter essentially guides first the consecutive population of $k_0=0$, $k_0/2=1.57$ and $k_0=3.14$ and subsequently of the 
exactly opposite momenta.   
This process repeats during the evolution. 
The different quench amplitudes impact mainly the speed of the alternating 
activation of momenta and to a lesser extent the magnitude of their population, 
compare Figs. \ref{fig:largersystem} ($o$) and ($m$).

\section*{Appendix B: The Computational Quantum Dynamics Approach MCTDHB} 

To simulate the nonequilibrium dynamics and calculate the stationary properties of the spatially interacting bosons we 
solve the many-body Schr\"{o}dinger equation $\left( {i\hbar {\partial _t} - H} \right)\ket{\Psi (t)} = 0$, by  
employing the Multi-Configuration Time-Dependent Hartree method for Bosons (MCTDHB) \cite{MCTDHB1,MCTDHB2,Streltsov}. 
This method has been applied extensively in several nonequilibrium bosonic settings, 
see e.g. \cite{Streltsov,Streltsov1,Alon2,Alon3,Jannis,Mistakidis,Mistakidis1,Mistakidis7,Koutentakis,Mistakidis4}. 
We remark that within our implementation we use the Multi-Layer Multi-Configuration 
Time-Dependent Hartree method for bosonic and fermionic Mixtures (ML-MCTDHX) \cite{Cao_X,Kronke_ML,Katsimiga_mixture,Katsimiga_mixture1}.  
The latter consists an extended version of the MCTDHB and is particularly suitable for treating multicomponent ultracold systems,   
while for the case of a single bosonic species it reduces to MCTDHB. 
MCTDHB is based on the usage of a time-dependent (t-d) and variationally optimized many-body basis set, 
which enables for the optimal truncation of the total Hilbert space. 
The expansion of the many-body wavefunction relies on a linear combination of t-d permanents $\ket{\vec{n}}$ and t-d weights $A_{\vec{n}}(t)$  
\begin{equation}
\label{eq:10}\left| {\Psi (t)} \right\rangle  = \sum\limits_{\vec n
} {{A_{\vec n }}(t)\ket{\vec{n}} }.
\end{equation}
The bosonic number states $\ket{\vec{n}}=\left| {{n_1},{n_2},...,{n_M};t}\right\rangle$, built upon t-d 
single-particle functions (SPFs) $\left| \phi_{i}(t) \right\rangle$, $i=1,2,...,M$, with $M$ being the number of the considered SPFs. 
The summation $\vec n$ is taken over all the possible combinations $n_{i}$ such that the total number of bosons $N$ is conserved. 
Moreover, the SPFs are expanded within a time-independent primitive basis $\{\ket{k}\}$ of dimension $M_{pr}$. 
Within our implementation a $\rm{sine}$ discrete variable representation has been used as a primitive basis for the SPFs.  
We remark here that in the case of $M=1$ the many-body wavefunction is given by a single permanent $\ket{n_{1}=N;t}$ and the method 
reduces to the t-d Gross Pitaevskii mean-field approximation. 
\begin{figure}[ht]  
  \centering
\includegraphics[width=0.45\textwidth]{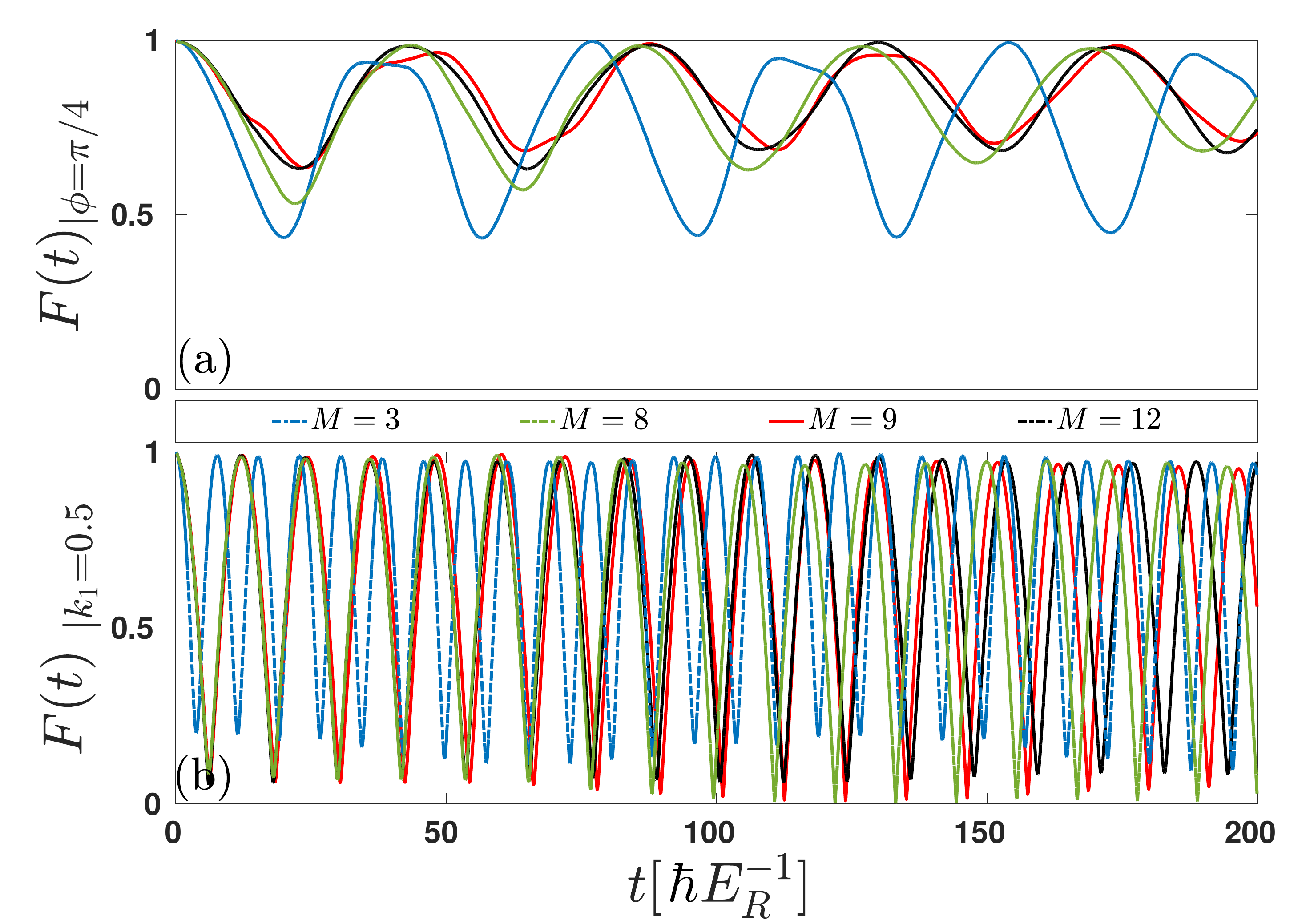}
  \caption{$F(t)$ for different number of SPFs (see legend) following a quench on ($a$) the phase $ \phi$ from $\phi=0$ to $\phi=\pi/4$   
  and ($b$) the wavevector $k_1$ from $k_1=0$ to $k_1=0.5$.  } 
  \label{fig:conver}
\end{figure} 

To obtain the t-d $N$-body wavefunction $\left|\Psi(t) \right\rangle$ under the influence of the Hamiltonian $\hat{H}$ we determine the equations 
of motion \cite{MCTDHB1,MCTDHB2,Streltsov} for the coefficients ${{A_{\vec n }}(t)}$ and the SPFs $\left| \phi_{i}(t) \right\rangle$ following e.g. the Dirac-Frenkel 
\cite{Frenkel,Dirac} variational principle, ${\bra{\delta\Psi}}{i{\partial _t} - \hat{ H}\ket{\Psi }}=0$. 
These equations consist of $\frac{(N+M-1)!}{N!(M-1)!}$ linear equations of motion for ${{A_{\vec n }}(t)}$ 
being coupled to the $M$ non-linear integrodifferential equations of motion for the SPFs.
To prepare the system in the ground state of the Hamiltonian $\hat{H}$ we utilize the so-called improved relaxation scheme \cite{Koutentakis}, which is briefly outlined below.       
Namely, we employ a certain number of SPFs $\lbrace |\phi_i^{(0)} \rangle \rbrace$ and  
diagonalize the Hamiltonian within the basis spanned by the SPFs.  
Setting the $n$-th obtained eigenvector as the $A_{\vec{n}}^{(0)}$-vector we propagate the SPFs in imaginary time within 
a finite time interval $d \tau$ and update the SPFs to $\lbrace |\phi_i^{(1)} \rangle \rbrace$. 
The above-mentioned steps are repeated until the energy of the state converges within the prescribed accuracy. 

To accurately perform the numerical integration of the MCTDHB equations of motion the overlap criteria 
$|\langle \Psi |\Psi \rangle -1| < 10^{-9}$ for the total wavefunction and
the SPFs $|\langle \varphi_i |\varphi_j \rangle -\delta_{ij}| < 10^{-10}$ are imposed. 
Moreover, we increase the number of variationally optimized SPFs and primitive basis states observing a systematic convergence of our results. 
For instance, we have used $M=9$, $M_{pr}=300$ for the triple well and $M=10$, $M_{pr}=400$ for the five well respectively. 
Let us next briefly demonstrate the convergence behaviour of our triple well simulations for an increasing number of SPFs.  
To achieve the latter we employ the extensively used, here, fidelity evolution upon quenching  
either the wavevector or the phase of the spatially-dependent interaction strength $C_{int}$.  
Fig. \ref{fig:conver} presents $F(t)$ for a sudden phase shift from $\phi=0$ to $\phi=\pi/4$ [see Fig. \ref{fig:conver} ($a$)] 
and a wavevector quench from $k_1=0$ to $k_1=0.5$ [see Fig. \ref{fig:conver} ($b$)] for different number of SPFs.
For reasons of completeness we remark that these quench amplitudes refer to enhanced response regions of the respective quench protocol. 
In both cases a systematic convergence of $F(t)$ is showcased for an increasing number of SPFs and in particular for $M>8$. 
Indeed following a phase quench, see Fig. \ref{fig:conver} ($a$), the maximum deviation observed in $F(t)$ 
between the 9 and 12 SPF cases is of the order of $8\%$.  
Turning to the wavevector quench, presented in Fig. \ref{fig:conver} ($b$), an admittedly better degree of convergence 
is observed throughout the evolution as the relative difference of $F(t)$ between 9 and 12 SPFs lies below $5\%$ and becomes at most $9\%$ for long 
propagation times $t>160$. 
An auxilliary indicator for the obtained numerical accuracy is provided by the population of the lowest occupied natural orbital which is   
kept below $0.1\%$ (not shown here for brevity).

\section*{Acknowledgements} 
 
S. I. M and P. S gratefully acknowledge financial support by the Deutsche Forschungsgemeinschaft 
(DFG) in the framework of the SFB 925 ``Light induced dynamics and control of correlated quantum
systems''. 
S. I. M thanks G. C. Katsimiga and G. M. Koutentakis for fruitful discussions.  

{}

\end{document}